\documentclass[journal,a4paper]{IEEEtran}
\usepackage{amsmath,amssymb,mathtools}
\usepackage{graphicx,booktabs,array,microtype,xcolor}
\usepackage{tikz}
\usetikzlibrary{arrows.meta,calc}
\usepackage{stfloats}
\usepackage{balance}
\usepackage{needspace}
\usepackage{placeins}
\usepackage[nocompress]{cite}
\usepackage{hyperref}
\definecolor{linkblue}{RGB}{25,70,120}
\hypersetup{colorlinks=true,linkcolor=linkblue,citecolor=linkblue,
 urlcolor=linkblue,pdftitle={Sequences for Flat-Top Beams}}
\allowdisplaybreaks[2]
\newtheorem{theorem}{Theorem}
\newtheorem{proposition}[theorem]{Proposition}
\newtheorem{lemma}[theorem]{Lemma}
\newtheorem{corollary}[theorem]{Corollary}
\newtheorem{remark}{Remark}
\newtheorem{definition}{Definition}
\newtheorem{Example}{Example}
\newcommand{\dd}{\,\mathrm d}
\newcommand{\e}{\mathrm e}
\newcommand{\sinc}{\operatorname{sinc}}
\newcommand{\dist}{\operatorname{dist}}

\newcommand{\Z}{\mathbb Z}
\newcommand{\ind}{\mathbf 1}
\newcommand\blfootnote[1]{%
  \begingroup
  \renewcommand\thefootnote{}\footnote{#1}%
  \addtocounter{footnote}{-1}%
  \endgroup
}
\title{Sequences for Flat-Top Beams}
\author{Cheng Du, \textit{Member, IEEE}}
\begin{document}
\maketitle
\blfootnote{Cheng Du is with the School of Computing and Information Technology, Great Bay University, Dongguan, China (E-mail: cdu@gbu.edu.cn).}
\begin{abstract}
Flat-top beams have numerous applications in radar and communication systems. Sequences with diverging merit factors can approach the ideal omnidirectional beams but do not directly apply to flat-top
beams, which require simultaneous control of passband ripple
and stopband leakage. To bridge this gap, we study the asymptotic behavior of both quantities of flat-top beams by using tools of Fourier analysis. We establish contrasting results for sequences with finite or infinite alphabets. For sequences with finite alphabet, including binary and quaternary ones, their passband ripple and stopband leakage cannot simultaneously vanish as the sequence length grows to infinity. For sequences with infinite alphabet, however, both requirements are satisfied simultaneously by the generalized step-chirp (GSC) sequences if and only if the length-bandwidth product $N\gamma$ goes to infinity and the distance of the step size $a$ from $\{0,1,\tfrac12,\tfrac13,\ldots\}$ goes to zero. Besides, both quantities vanish at rate
\(O((N\gamma)^{-1/2})\) if $a\in \{1,\tfrac12,\tfrac13,\ldots\}$. 
These results broaden the scope of classical sequence design beyond full-band spectral flatness and provide a theoretical foundation for flat-top beam synthesis.
\end{abstract}

\section{Introduction}
Flat-top beam synthesis is a classical problem in antenna array
design~\cite{Ksienski1960,OrchardElliottStern1985}: choosing excitation weights to obtain nearly uniform power
over a prescribed angular sector and low radiation outside
it. Such beams are useful for radar sector search, where targets must
be illuminated across a surveillance
region~\cite{KerceBrownMitchell2007}, and for beam sweeping in cellular communication systems to send common messages to user equipment (UE) at the initial access phase~\cite{GirnykPetersson2021,DuJiang2024}. Flat-top beams can also improve detection probability during beam
search in massive multiple-input multiple-output (MIMO)  communication systems~\cite{XiaoXiaXia2017}.

In contrast to flat-top beams, omnidirectional beams spread
energy nearly uniformly over the entire angular range. In cellular communication systems, they
are used to broadcast common messages  without beam sweeping~\cite{XiaGao2016,GirnykPetersson2021,LiJiangDuWang2021}.
This cell-wide broadcast requirement is specified by
3GPP~\cite[Sec.~5.5]{3GPPTS38300}, and commercial 5G radios
already employ a single broad beam for synchronization signal
block transmission~\cite{GirnykJidhageFaxer2023}.

For uniform linear arrays (ULAs),
the beam patterns are essentially the \emph{continuous-frequency} power
spectra of the excitation vectors~\cite{DuJiang2024,LiJiangDuWang2021}, thus determined by their \emph{aperiodic} autocorrelations by the Wiener--Khinchin theorem. This motivates studying beam synthesis through
sequence design. Specifically, for omnidirectional beams, asymptotically vanishing full-band ripple requires an excitation sequence with its merit factor tending to
infinity~\cite{Golay1982,Schmidt2016Survey}; for flat-top beams, \textit{passband ripple} is governed by a quadratic form in the aperiodic autocorrelations 
while \textit{stopband leakage} is an affine function of the
autocorrelations, as detailed in Subsection~\ref{m:isl-connection}.

A second motivation is in the aspect of practical hardware implementation.
Polyphase sequences, widely studied in sequence
design~\cite{Schmidt2016Survey}, yield equal-amplitude excitations
that can be implemented with a single radio-frequency (RF) chain
and a phase-shifter network (PSN)~\cite{XiaoXiaXia2017}.
This reduces the number of costly digital-to-analog converters
(DACs) at the transmitter and analog-to-digital converters (ADCs)
at the receiver compared with fully digital
arrays~\cite{HeathEtAl2016}. Besides, unimodular excitations maintain uniform loading of the power amplifiers (PAs), thereby avoiding the additional power back-off and potentially improving
PA energy efficiency~\cite{DuJiang2024,GirnykPetersson2021}. Moreover, binary and quaternary alphabets, common in sequence
design~\cite{Golay1961,DuJiangGolay2024}, require only one- and
two-bit phase shifters, respectively, further reducing the cost
of the PSN~\cite{HeathEtAl2016}.

\subsection{Related work}
\emph{Phase-only beam synthesis.}
Phase-only beam synthesis shapes the beam pattern by adjusting the
excitation phases while keeping their magnitudes fixed.
Optimization-based methods address the resulting nonconvex problem
under unit-modulus
constraints~\cite{CaoThompsonHaas2017,TranterEtAl2017,AroraEtAl2022}. Analytical approaches use nonlinear-FM
tapers~\cite{ColemanEtAl2010}, quadratic-phase (sampled-LFM)
excitations~\cite{FonteneauEtAl2022}, or transport
maps~\cite{AngelettiEtAl2026I,AngelettiEtAl2026II}. In particular,
Generalized Step-Chirp (GSC) sequences extend sampled LFM to
accommodate low-bit phase shifters, with
numerical evidence suggesting asymptotically vanishing passband
ripple and stopband leakage for certain parameter
choices~\cite{DuJiang2024}.

These approaches control the beamwidth of flat-top beams flexibly through
numerical optimization or analytical parameter selection. However, rigorous
analysis of the asymptotic behavior of passband ripple and stopband leakage as the number of antennas grows remains
limited.

\emph{High-merit-factor sequences.}
Littlewood conjectured that the optimal merit factor of binary sequences can
be arbitrarily large~\cite{Littlewood1966}, whereas Golay's heuristic
analysis predicted a finite asymptotic optimum of
\(12.3248\ldots\)~\cite{Golay1982}.
Explicit constructions give lower bounds on the optimal
asymptotic merit factor: Rudin--Shapiro-type recursions yield
sequences with limiting merit factor
\(3\)~\cite{HoholdtJensenJustesen1985,KatzLeeTrunov2020}.
Constructions based on difference sets and finite-field
characters~\cite{JensenJensenHoholdt1991,Katz2013L4} include appended
rotated Legendre sequences with limiting merit factor
\(6.342061\ldots\)~\cite{BorweinChoiJedwab2004,JedwabKatzSchmidt2013}.
Numerical optimization~\cite{Baden2011} and studies of demerit-factor
distributions~\cite{KatzRamirez2025} complement these constructions.
Remarkably, a
recent preprint by OpenAI claims the existence of binary sequences with
diverging merit factor~\cite[Cor.~8.1]{OpenAI2026Littlewood}, which, if correct, would prove Littlewood's conjecture.

When the phase alphabet is allowed to grow with sequence length, some
polyphase sequences have long been known to achieve diverging merit factors.
Classical examples include Frank and Chu
sequences~\cite{Frank1963,Chu1972}, whose merit factors are
asymptotic to \((\pi^2/4)\sqrt N\) and \((\pi/2)\sqrt N\),
respectively~\cite{Mercer2013,Schmidt2013}.
The piecewise linear polyphase codes of Petrolati, Angeletti, and Toso
(PAT codes) provide further constructions with favorable finite-length
merit factors~\cite{PetrolatiAngelettiToso2012}.
With arbitrary phases, numerical integrated sidelobe level (ISL)
minimization produces additional
sequence designs~\cite{BorweinFerguson2005,StoicaHeLi2009,SongBabuPalomar2015},
and ultraflat-polynomial constructions establish uniform convergence
of the normalized power spectrum to a constant, which also implies
diverging merit factors~\cite{Kahane1980,BombieriBourgain2009}.

While these sequence-design results provide a theoretical basis
for omnidirectional beam synthesis through full-band spectral flatness,
they do not directly extend to flat-top beam synthesis,
where power must be concentrated within a prescribed sector.

\addtolength{\dbltextfloatsep}{-12pt}
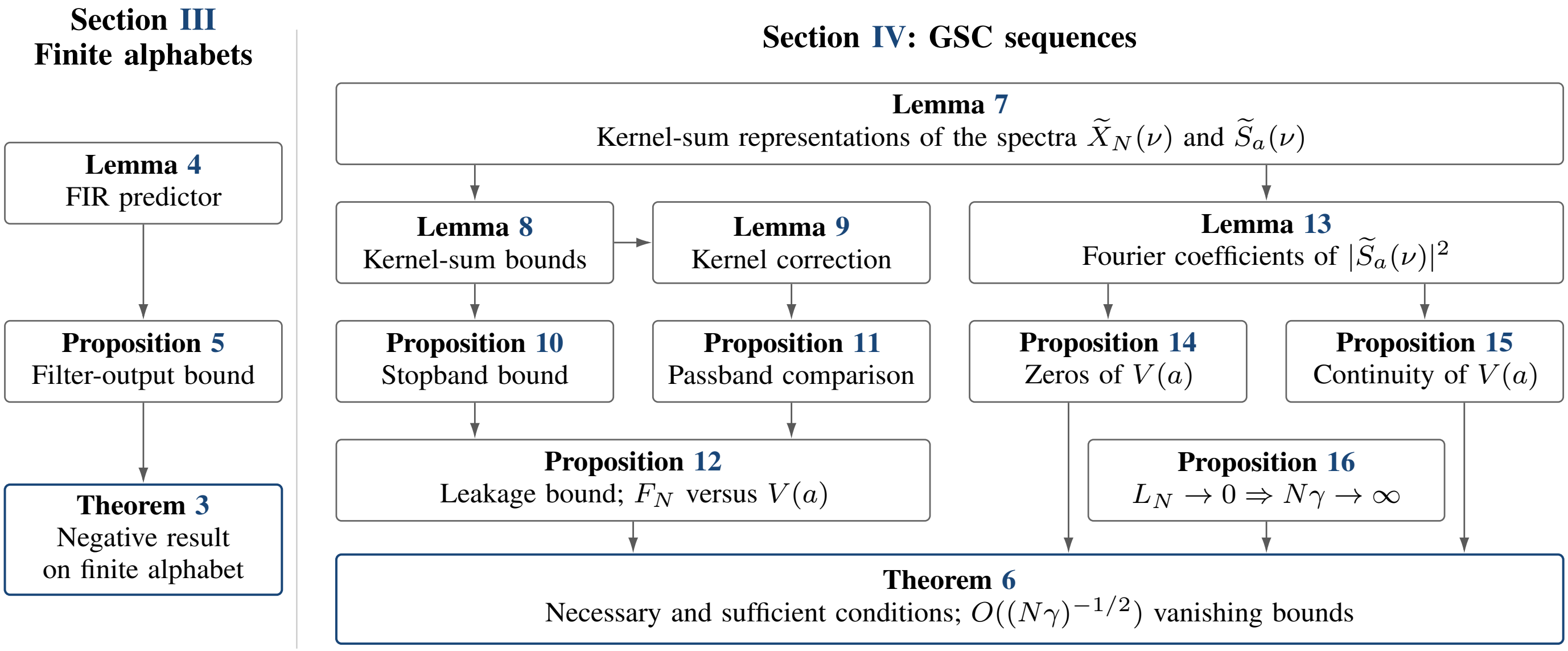
\begin{figure*}[!b]
\centering
\resizebox{\textwidth}{!}{%
\begin{tikzpicture}[
  x=1cm,y=1cm,
  result/.style={draw=black!60,fill=white,rounded corners=1.5pt,
    line width=.45pt,align=center,text width=2.6cm,minimum width=2.8cm,
    minimum height=.82cm,inner xsep=.1cm,inner ysep=3pt,
    font=\fontsize{8}{9.2}\selectfont},
  conclusion/.style={result,draw=linkblue,line width=.65pt},
  flow/.style={draw=black!65,line width=.45pt},
  edge/.style={flow,-{Latex[length=1.6mm,width=1mm]}},
  heading/.style={font=\small\bfseries,align=center}
]
\node[heading] at (1.45,.85) {Section~\ref{fa:section}\\Finite alphabets};
\node[heading] at (9.6,.85) {Section~\ref{m:construction}: GSC sequences};
\draw[black!20] (3,.95)--(3,-5.3);

\node[result] (l2) at (1.45,-.6)
  {\textbf{Lemma~\ref{fa:lem-stable-filter}}\\FIR predictor};
\node[result] (p3) at (1.45,-2.4)
  {\textbf{Proposition~\ref{fa:prop-fixed-filter}}\\Filter-output bound};
\node[conclusion] (t1) at (1.45,-4.2)
  {\textbf{Theorem~\ref{fa:thm-obstruction}}\\Negative result\\on finite alphabet};
\draw[edge] (l2)--(p3);
\draw[edge] (p3)--(t1);

\node[result,text width=12.2cm,minimum width=12.4cm] (p5) at (9.6,0)
  {\textbf{Lemma~\ref{m:prop-fourier-representations}}\\Kernel-sum representations of the spectra $\widetilde{X}_N(\nu)$ and $\widetilde{S}_a(\nu)$};
\node[result] (l6) at (4.8,-1.2)
  {\textbf{Lemma~\ref{z:lem-abel-tail}}\\Kernel-sum bounds};
\node[result] (l7) at (8,-1.2)
  {\textbf{Lemma~\ref{z:lem-correction-weight}}\\Kernel correction};
\node[result,text width=5.8cm,minimum width=6cm] (l11) at (12.8,-1.2)
  {\textbf{Lemma~\ref{z:prop-fourier-coefficients}}\\Fourier coefficients of $|\widetilde{S}_a(\nu)|^2$};
\node[result] (p8) at (4.8,-2.4)
  {\textbf{Proposition~\ref{z:prop-stopband}}\\Stopband bound};
\node[result] (p9) at (8,-2.4)
  {\textbf{Proposition~\ref{z:prop-cellwise}}\\Passband comparison};
\node[result] (p12) at (11.2,-2.4)
  {\textbf{Proposition~\ref{z:coro-V}}\\Zeros of \(V(a)\)};
\node[result] (p13) at (14.4,-2.4)
  {\textbf{Proposition~\ref{z:prop-V-continuity}}\\Continuity of \(V(a)\)};
\node[result,text width=5.8cm,minimum width=6cm] (p10) at (6.4,-3.6)
  {\textbf{Proposition~\ref{z:prop-uniform-comparison}}\\Leakage bound; \(F_N\) versus \(V(a)\)};
\node[result,text width=3.4cm,minimum width=3.6cm] (p14) at (12.8,-3.6)
  {\textbf{Proposition~\ref{z:prop-dpss}}\\\mbox{\(L_N\to0\Rightarrow N\gamma\to\infty\)}};
\node[conclusion,text width=12.2cm,minimum width=12.4cm,minimum height=.9cm]
  (t4) at (9.6,-4.8)
  {\textbf{Theorem~\ref{thm:vanishing}}\\Necessary and sufficient conditions;
   \(O((N\gamma)^{-1/2})\) vanishing bounds};

\draw[edge] ([xshift=3.2cm]p5.south)--(l11.north);
\draw[edge] ([xshift=-4.8cm]p5.south)--(l6.north);
\draw[edge] (l6.east)--(l7.west);
\draw[edge] (l6.south)--(p8.north);
\draw[edge] (l7.south)--(p9.north);
\draw[edge] (p8.south)--([xshift=-1.6cm]p10.north);
\draw[edge] (p9.south)--([xshift=1.6cm]p10.north);

\draw[edge] ([xshift=-1.6cm]l11.south)--(p12.north);
\draw[edge] ([xshift=1.6cm]l11.south)--(p13.north);
\draw[edge] (p10.south)--([xshift=-3.2cm]t4.north);
\draw[edge] ([xshift=-.4cm]p12.south)--([xshift=1.2cm]t4.north);
\draw[edge] ([xshift=.4cm]p13.south)--([xshift=5.2cm]t4.north);
\draw[edge] (p14.south)--([xshift=3.2cm]t4.north);
\end{tikzpicture}%
}
\caption{Organization of Sections~\ref{fa:section}
 and~\ref{m:construction}.}
\label{fig:proof-structure}
\end{figure*}

\makeatletter
\AddToHookNext{cmd/@cdblflb/before}{%
  \setbox\@outputbox\vbox{\box\@outputbox}%
}
\makeatother
\balance
\subsection{Question and contributions}
\label{m:alphabet-constraints}
To bridge the gap between flat-top beam synthesis and sequence design,
we study whether the passband ripple and stopband leakage of flat-top beams can vanish simultaneously as the sequence length goes to infinity, i.e., whether the ideal flat-top beam can be approached asymptotically.
It turns out that the answer depends on the alphabet size.

\begin{enumerate}
 \item \emph{Negative result on finite alphabet.}
 We prove that as the sequence length goes to infinity, the passband ripple and stopband leakage cannot vanish simultaneously for any finite complex alphabet, including the binary and quaternary ones, which would otherwise
be particularly attractive for implementation with a low-cost PSN. This contrasts with the full-band counterpart, namely the recent claim of binary sequences with
 diverging merit factors~\cite{OpenAI2026Littlewood}.

 \item \emph{Positive result on infinite alphabet.}
We prove that the passband ripple and the stopband leakage of a GSC sequence, with an infinite alphabet, vanish simultaneously if and only if $N\gamma\to\infty$ and $\dist(a,\{0,1,\tfrac12,\tfrac13,\ldots\})\to0$, where $N$ is the sequence length, $\gamma$ is the normalized beamwidth, and $a$ is the step size of the GSC sequence. Moreover, both quantities converge at rates $O((N\gamma)^{-1/2})$ if $a \in\{1,\tfrac12,\tfrac13,\ldots\}$, which guarantees the asymptotic performance of flat-top beams.
\end{enumerate}

The remainder of the paper is organized as follows.
Section~\ref{m:problem-formulation} presents the preliminaries, including the beam synthesis model and the spectral criteria, along with anautocorrelation interpretation.
Section~\ref{fa:section} establishes the negative result on finite alphabets:
Lemma~\ref{fa:lem-stable-filter} enables the filter construction in
Proposition~\ref{fa:prop-fixed-filter}, whose estimate yields
the main result, Theorem~\ref{fa:thm-obstruction}.
Section~\ref{m:construction} proves the positive result on GSC sequences. The basic idea is to compare the spectra of GSC sequences $\widetilde{X}_N(\nu)$ with those of continuous step-chirp
signals $\widetilde{S}_a(\nu)$. The resulting bounds control the stopband leakage $L_N$ and relate
the passband ripple $F_N$ to a limiting function \(V(a)\). The zeros and
continuity of $V(a)$, combined with the necessary condition
\(N\gamma\to\infty\), yield the parameter conditions for simultaneous
vanishing of $F_N$ and $L_N$; the same bounds give the convergence rates for $a \in \{1,\tfrac12,\tfrac13,\ldots\}$. Figure~\ref{fig:proof-structure} summarizes the proof structure
of Sections~\ref{fa:section} and~\ref{m:construction}. Some numerical illustrations are also provided in this section to demonstrate the theoretical findings. The final section discusses future directions, and the appendices
contain auxiliary proofs.

\emph{Notation.}
The sets of integers, positive integers, real numbers, and complex
numbers are denoted by \(\Z\), \(\Z^+\), \(\mathbb R\), and
\(\mathbb C\), respectively, and \(\Z_M=\{0,\ldots,M-1\}\) for
\(M\in\Z^+\). For a finite set \(B\), \(|B|\) denotes its cardinality;
for a frequency band \(\mathcal B\), \(|\mathcal B|\) denotes its Lebesgue
measure. Boldface denotes sequences, vectors, and matrices,
and \(\|\cdot\|_2\) is the Euclidean norm.
We write \(i\) for the imaginary unit, \(\overline z\) for the
complex conjugate of \(z\), \(\Re z\) for its real part, and
\((\cdot)^*\) for the conjugate transpose.
The indicator of a set \(E\) is \(\ind_E\), and
\(\dist(t,E)=\inf_{s\in E}|t-s|\) for nonempty \(E\subseteq\mathbb R\).
The floor \(\lfloor t\rfloor\) is the greatest integer not exceeding
\(t\). We use the normalized sinc function
\(\sinc t=\sin(\pi t)/(\pi t)\), with \(\sinc0=1\).
The notation \(u=O(v)\) means that \(|u|\leq C|v|\) for some constant
\(C\).

\clearpage
\addtolength{\dbltextfloatsep}{12pt}
\nobalance
\section{Preliminaries}
\label{m:problem-formulation}

\subsection{Beam synthesis and spectral criteria}
\label{m:ula-flat-top}
\begin{figure}[!ht]
 \centering
 \includegraphics[width=\columnwidth]{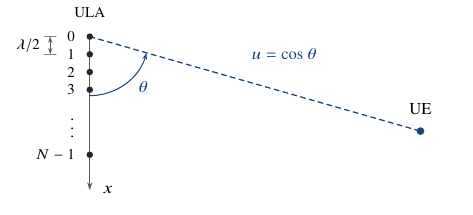}
 \caption{ULA--UE geometry.}
 \label{fig:ula-geometry}
\end{figure}

Consider the ULA in Fig.~\ref{fig:ula-geometry}, with
\(N\) isotropic antennas along the \(x\)-axis with spacing
\(\lambda/2\), where \(\lambda\) is the carrier
wavelength. For a UE at azimuth angle \(\theta\) in the far field, antenna \(n\) contributes
a relative propagation phase \(\e^{-i\pi n \cos\theta}\). When the excitations
\(\mathbf x=(x_0,\ldots,x_{N-1})\in\mathbb C^N\) are applied to the transmit antennas, the normalized received signal is the sum of the contributions from all antennas, as given by
\begin{equation}
 X_N(f)\triangleq\frac1{\|\mathbf x\|_2}\sum_{n=0}^{N-1}x_n\e^{-2\pi inf},\ f=\frac{1}{2}\cos\theta\pmod1.
 \label{m:spectrum}
\end{equation}
 The angular distribution of received power $|X_N(f)|^2$ for $\theta \in [0,\pi]$ is referred to as the beam pattern. Beam synthesis is therefore a problem of shaping the power spectrum
\(|X_N(f)|^2\). 

Multiple UEs may reside within an angular range, and the beam should provide nearly uniform gain for them while minimizing leakage outside the angular range. This leads to the following definitions of passband ripple and stopband leakage.

\begin{definition}
\label{m:def-omni-flat-top}
Let $\mathcal B \subseteq [0,1)$ be a prescribed interval. The passband ripple and the stopband leakage of the beam pattern $|X_N(f)|^2$ are defined as
\begin{align}
 F_N&\triangleq\frac{\gamma}{(1-L_N)^2}
 \int_{\mathcal B}\left(|X_N(f)|^2-\frac{1-L_N}{\gamma}\right)^2\dd f,
 \label{m:ripple}\\
 L_N&\triangleq\int_{\mathcal B^c}|X_N(f)|^2\dd f,\label{m:leakage}
\end{align}
respectively, where $\mathcal B^c$ is the complement of $\mathcal B$ in $[0,1)$ and $\gamma = |\mathcal B|$ is the beamwidth.
\end{definition}

It is also natural to use the passband error relative to the ideal level \(1/\gamma\):
\begin{equation}
 D_N\triangleq\gamma\int_{\mathcal B}
       \left(|X_N(f)|^2-\frac1\gamma\right)^2\dd f.
 \label{m:D}
\end{equation}
It is straightforward to verify that 
\begin{equation}
D_N=(1-L_N)^2F_N+L_N^2.
 \label{m:centering}
\end{equation}

\begin{definition}
      Given small thresholds $\epsilon_1, \epsilon_2 \geq 0$, the beam pattern $|X_N(f)|^2$ is omnidirectional if $\mathcal B = [0,1)$ and $F_N \leq \epsilon_1$, or flat-top if $\mathcal B \neq [0,1)$ and $F_N \leq \epsilon_1, L_N\leq \epsilon_2$.
\end{definition}

\begin{figure}[!ht]
 \centering
 \includegraphics[width=\columnwidth]{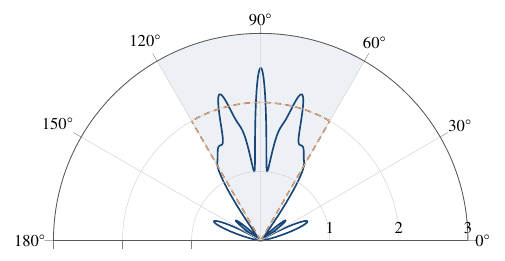}
 \caption{A flat-top beam for \(N=16\) and \(\gamma=1/2\).}
 \label{fig:qpsk16-beam}
\end{figure}

\begin{Example}
Consider \(N=16\) and
\(\mathcal B=[-\tfrac14,\tfrac14)\pmod1\), corresponding to
\(60^\circ\leq\theta\leq120^\circ\). Let
\[
 \mathbf x=[1,i,i,-1,-i,1,1,1,1,1,1,-i,-1,i,i,1]^\top.
\]
Figure~\ref{fig:qpsk16-beam} shows the flat-top beam synthesized by the sequence \(\mathbf x\) (solid
curve) with \(F_{16}\approx0.0497\) and \(L_{16}\approx0.1782\). The dashed curve marks the ideal gain \(2\), and shading indicates
the target angular range.
\end{Example}

The passband ripple and the stopband leakage are crucial for applications of flat-top beams. For example, for beam search in cellular communications, the average detection performance over the target angular range attains its ideal upper bound if and only if $\epsilon_1 = \epsilon_2 = 0$~\cite[Property~3]{XiaoXiaXia2017}. However, this is unattainable since $\mathbf x$ is a finite impulse response (FIR) filter which cannot vanish on a band. Thus our goal is to investigate whether \(\epsilon_1>0\) and \(\epsilon_2>0\) can be arbitrarily small for sufficiently large $N$, i.e., whether $F_N, L_N \to 0$ as $N \to \infty$.

\subsection{Hardware considerations for alphabet selection}
\label{m:hardware-implementation}
\begin{figure}[!ht]
 \centering
 \includegraphics[width=\columnwidth]{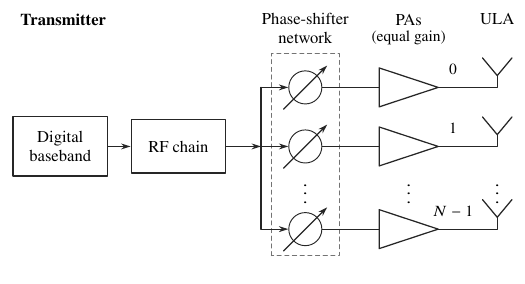}
 \caption{Hardware implementation of beam synthesis based on unimodular sequences.}
\label{fig:ula-transceiver}
\end{figure}

For a sequence \(\mathbf x\in \mathcal U^N\) with \(\mathcal U \subset \mathbb C\), $\mathcal U$ is referred to as the alphabet, which is referred to be finite (infinite) if it contains only finitely (infinite) many elements. The hardware implementation of beam synthesis depends on the alphabet of the excitation sequence. If the alphabet is non-unimodular, i.e., there exist $\alpha, \alpha' \in \mathcal U$ such that $|\alpha|\ne|\alpha'|$, then it requires multiple RF chains which could be costly for large arrays. However, if the alphabet is unimodular, i.e., $|\alpha|=|\alpha'|$ for any $\alpha, \alpha' \in \mathcal U$, it allows implementation with a single RF chain shared
across the antennas, thereby lowering the hardware cost of RF chains. Figure~\ref{fig:ula-transceiver} illustrates such an implementation, where a digital baseband signal is first converted to an RF signal, which is then split equally among \(N\) branches with phase shifters and equal-gain PAs.

A small alphabet \(\mathcal U\) further reduces the
number of phase states that the PSN must realize. For example, a binary
alphabet \(\mathcal U = \{\pm1\}\) requires one-bit phase shifters, while a quaternary alphabet
\(\mathcal U = \{\pm1,\pm i\}\) requires
two-bit phase shifters. Thus a small alphabet is desirable for low-cost hardware implementation of the PSN.

\subsection{Connection to Aperiodic Autocorrelation}
\label{m:isl-connection}
For \(\mathbf x\in \mathbb C^N\), its aperiodic
autocorrelation is defined as 
\begin{equation}
 r_{k}\triangleq\sum_{n=0}^{N-1-k}x_{n+k}\overline{x_n},
 \quad k=0,\ldots,N-1,
 \label{m:autocorrelation-isl}
\end{equation}
and \(r_{-k}=\overline{r_k}\)
for \(k=1,\ldots,N-1\). The merit factor of $\mathbf x$ is defined as
\begin{equation}
      \operatorname{MF}(\mathbf x) \triangleq \frac{\|\mathbf x\|_2^4}{\sum_{0<|k|<N}|r_k|^2}.
\end{equation}

Let \(\mathcal B=[f_0-\gamma/2,f_0+\gamma/2)\pmod1\) and define 
\begin{equation}
 \beta_k\triangleq \gamma\e^{2\pi ikf_0}\sinc(\gamma k),\qquad k\in\Z.
\end{equation}
The following proposition relates the passband ripple and stopband leakage to the aperiodic autocorrelation of \(\mathbf x\).
\begin{proposition}
\label{m:prop-correlation-criteria}
The stopband leakage $L_N$ and the passband ripple $F_N$ satisfy
\begin{equation}  \label{m:ripple-correlation}
 L_N=1-\frac{1}{\|\mathbf x\|_2^2} \boldsymbol\beta^*\mathbf r, \qquad
 F_N=\frac{\mathbf r^*\mathbf W\mathbf r}
                    {\|\mathbf x\|_2^4(1-L_N)^2},
\end{equation}
where  
\begin{equation}
      \begin{aligned}
                 \mathbf r=&\ (r_{1-N}, \ldots, r_{-1}, r_0, r_1, \ldots, r_{N-1})^\top,\\
      \boldsymbol\beta=&\ (\beta_{1-N}, \ldots, \beta_{-1}, \beta_0, \beta_1, \ldots, \beta_{N-1})^\top,\\
 \mathbf W
 =&\ \gamma\operatorname{Toep}(\beta_0,\beta_1,\ldots,\beta_{2N-2})-\boldsymbol\beta\boldsymbol\beta^*,
       \end{aligned}
 \label{m:ripple-matrix}
\end{equation}
with \(\operatorname{Toep}(\beta_0,\beta_1,\ldots,\beta_{2N-2})\) representing the
Hermitian Toeplitz matrix with first column
\((\beta_0,\beta_1,\ldots,\beta_{2N-2})^\top\).
\end{proposition}

\begin{IEEEproof}
The Wiener--Khinchin theorem gives
\begin{equation}
 |X_N(f)|^2=\frac1{\|\mathbf x\|_2^2}
       \sum_{|k|<N}r_k\e^{-2\pi ikf}.
 \label{m:acf-spectrum}
\end{equation}
Note that \(\beta_k=\int_{\mathcal B}\e^{2\pi ikf}\dd f\). Integrating
\eqref{m:acf-spectrum} over \(\mathcal B\) gives
\[
 1-L_N=\frac1{\|\mathbf x\|_2^2}\sum_{|k|<N}r_k\overline{\beta_k}
       =\frac{\boldsymbol\beta^*\mathbf r}{\|\mathbf x\|_2^2},
\]
which proves the leakage identity in \eqref{m:ripple-correlation}.
Besides,
\begin{equation}
|X_N(f)|^2-\frac{1-L_N}{\gamma} =\sum_{|k|<N} \frac{r_k}{\|\mathbf x\|_2^2}
       \left(\e^{-2\pi ikf}-\frac{\overline{\beta_k}}\gamma\right)
 \label{m:centered-acf-spectrum}
\end{equation}
Thus
\[
 \begin{aligned}
 &\gamma \|\mathbf x\|_2^4\int_{\mathcal B}
       \left(|X_N(f)|^2-\frac{1-L_N}{\gamma}\right)^2\dd f\\
 =&\ \gamma\sum_{|k|,|\ell|<N}\overline{r_k}r_\ell
       \int_{\mathcal B}
       \left(\e^{2\pi ikf}-\frac{\beta_k}{\gamma}\right) \cdot
       \left(\e^{-2\pi i\ell f}-\frac{\overline{\beta_\ell}}{\gamma}\right)\dd f\\
=&\ \sum_{|k|,|\ell|<N}\overline{r_k}
       \bigl(\gamma \beta_{k-\ell}-\beta_k\overline{\beta_\ell}\bigr)r_\ell =  \mathbf r^*\mathbf W\mathbf r,
 \end{aligned}
\]
which proves the ripple identity in \eqref{m:ripple-correlation} by checking the definition of $F_N$ in \eqref{m:ripple}.
\end{IEEEproof}

\begin{corollary}
\label{m:cor-full-band}
If \(\gamma=1\), then $F_N \to 0 \Longleftrightarrow \operatorname{MF}(\mathbf x) \to \infty$.
\end{corollary}
\begin{IEEEproof}
If \(\gamma=1\), then we have \(\beta_0=1\) and \(\beta_k=0\) for all \(k\ne0\).
Thus \(\boldsymbol\beta^*\mathbf r=r_0 = \|\mathbf x\|_2^2\) and
\[
 \mathbf W=\operatorname{diag}\!\left(
 \underbrace{1,\ldots,1}_{N-1},\,0,\,
 \underbrace{1,\ldots,1}_{N-1}
 \right).
\]
By \eqref{m:ripple-correlation}, $L_N = 0$ and $F_N=\frac{1}{\|\mathbf x\|_2^4}\sum_{0<|k|<N}|r_k|^2 = \frac{1}{\operatorname{MF}(\mathbf x)}$, thus $F_N \to 0 \Leftrightarrow \operatorname{MF}(\mathbf x) \to \infty$.
\end{IEEEproof}

The above corollary shows that the requirement of approaching an ideal omnidirectional beampattern is equivalent to designing sequences with diverging merit factors. However, the requirement of approaching an ideal flat-top beampattern, i.e., $\lim_{N\to\infty} L_N = 0$ and $\lim_{N\to\infty}F_N = 0$ for $\gamma < 1$, is beyond the scope of classical sequence design and will be investigated in the subsequent sections of this work.


\Needspace{17\baselineskip}
\section{Negative result on finite alphabet}
\label{fa:section}
Motivated by the advantages of a small unimodular alphabet in hardware implementation, we ask whether it is possible for a binary or quaternary sequence to have $\lim_{N\to \infty} F_N=0$ and $\lim_{N\to \infty} L_N=0$ simultaneously. The following theorem gives a negative answer for an even broader class of sequences.

\begin{theorem}
\label{fa:thm-obstruction}
If \(\mathcal B\subsetneq[0,1)\), then for any
\(\mathbf x\in\mathcal U^N\) with finite alphabet \(\mathcal U\),
the stopband leakage \(L_N\) and the passband ripple \(F_N\)
cannot both vanish as \(N\to\infty\).
\end{theorem}

\begin{remark}
      OpenAI's recent preprint~\cite{OpenAI2026Littlewood} shows the existence of binary sequences with $\lim_{N\to\infty}F_N=0$ if $\mathcal B = [0, 1)$, which means that the ideal omnidirectional beam can be approached by using one-bit phase shifters. In contrast, a direct corollary of Theorem~\ref{fa:thm-obstruction} rules out the existence of binary sequences with $\lim_{N\to\infty}F_N=0$ and $\lim_{N\to\infty}L_N=0$ simultaneously if $\mathcal B \subsetneq [0,1)$, which means that ideal \textit{flat-top} beams can never be approached by using one-bit phase shifters. Indeed, according to Theorem~\ref{fa:thm-obstruction}, it requires an infinite alphabet.
\end{remark}

To prove Theorem~\ref{fa:thm-obstruction}, we first construct an FIR filter \(p(z)\)
with small sub-band energy.

\begin{lemma}
\label{fa:lem-stable-filter}
For any sub-band \(\mathcal B\subsetneq[0,1)\) and any \(\varepsilon>0\), there exists
an FIR filter
\[
 p(z)=\sum_{k=0}^{m}p_kz^k,\qquad p_0=1,\quad m\geq1,
\]
depending only on \(\mathcal B\) and \(\varepsilon\), such that
\begin{equation}
 \sum_{k\in\Z}
 \left|\int_{\mathcal B}p(\e^{-2\pi if})\e^{2\pi ikf}\dd f\right|
 <\varepsilon.
 \label{fa:eq-small-filter}
\end{equation}
\end{lemma}

The proof is provided in Appendix~\ref{app:finite-alphabet}.

Based on the filter \(p(z)\) from Lemma~\ref{fa:lem-stable-filter}, we then construct another FIR filter \(q(z)\) whose output power for all finite-alphabet inputs is controlled by the stopband leakage.

\begin{proposition}
\label{fa:prop-fixed-filter}
There exist an FIR filter \(q(z)\) and a constant \(C>0\), both
depending only on the finite alphabet \(\mathcal U\) and the sub-band \(\mathcal B\subsetneq[0,1)\), such that for every nonzero \(\mathbf x\in\mathcal U^N\), 
\begin{equation}
 \int_0^1|q(\e^{-2\pi if})|^2|X_N(f)|^2\dd f
 \leq CL_N,\qquad \forall N\geq1,
 \label{fa:eq-fixed-filter-control}
\end{equation}
\begin{equation} \label{eq:qz_bound}
      |q(\e^{-2\pi if})|\leq1, \qquad \forall f\in[0,1).
\end{equation}
\end{proposition}


\begin{IEEEproof}
Let
\begin{equation}
 M\triangleq \max_{\alpha\in\mathcal U}|\alpha|,\qquad
 d\triangleq\min_{\substack{\alpha,\beta\in\mathcal U\\
                  \alpha\ne\beta}}|\alpha-\beta|>0.
 \label{fa:eq-alphabet-separation}
\end{equation}
Choose \(p(z)\) from Lemma~\ref{fa:lem-stable-filter}
with \(\varepsilon=d/(4M)\), and let \(Q\triangleq |\mathcal U|^m\). A block
\(\mathbf v=(v_{-m+1},\ldots,v_Q)\in\mathcal U^{m+Q}\) is called admissible if 
\begin{equation}
 \left|v_n+\sum_{k=1}^{m}p_k v_{n-k}\right|<\frac{d}{2},
 \qquad 1\leq n\leq Q.
 \label{fa:eq-admissible-block}
\end{equation}
For fixed \(v_{n-m},\ldots,v_{n-1}\) and \(1\leq n\leq Q\), two distinct candidates \(\alpha,\beta\in\mathcal U\) for \(v_n\) would satisfy
 \begin{equation}
 |\alpha-\beta|
 \leq\left|\alpha+\sum_{k=1}^{m}p_k v_{n-k}\right|
+\left|\beta+\sum_{k=1}^{m}p_k v_{n-k}\right|<d,
 \end{equation}
contradicting the fact that \(d\leq|\alpha-\beta|\). Therefore, an admissible block is uniquely determined by its initial state \((v_{-m+1},\ldots,v_0)\). Denote by \(\widehat Q\) the number of admissible blocks. Then \(\widehat Q\leq Q\). Form a matrix \(\mathbf V\in\mathcal U^{\widehat Q\times(Q+1)}\) by taking \((v_Q,v_{Q-1},\ldots,v_0)\) from these blocks as rows. Since the column vectors of \(\mathbf V\) are linearly dependent, there exist coefficients \(q_0,\ldots,q_Q\) such that \(\sum_{k=0}^{Q}|q_k|=1\) and, for every admissible block \(\mathbf v\), \(\sum_{k=0}^{Q}q_kv_{Q-k}=0\). Define the filter \(q(z)\triangleq\sum_{k=0}^{Q}q_kz^k\). Equation~\eqref{eq:qz_bound} follows from
\begin{equation}
       |q(\e^{-2\pi if})|
 \leq\sum_{k=0}^{Q}|q_k|=1.
\end{equation}

Take any \(N\geq1\) and a nonzero sequence \(\mathbf x=(x_0,\ldots,x_{N-1})\in\mathcal U^N\), and zero-extend \(\mathbf x\) to a bilateral sequence. Define the following two filter outputs:
\begin{equation}
 y_n\triangleq \sum_{k=0}^{m}p_kx_{n-k},\qquad w_n \triangleq\sum_{k=0}^{Q}q_kx_{n-k},\qquad n\in\Z.
 \label{fa:eq-pq-output}
\end{equation}
Given \(n\in\Z\), if \(|y_{n-k}|<d/2\) for all \(0\leq k<Q\), then the block \((x_{n-Q-m+1},\ldots,x_n)\) is admissible, and thus \(w_n=0\). The contrapositive is
\begin{equation}
      w_n\ne0
 \ \Longrightarrow\ \max_{0\leq k<Q}|y_{n-k}|\geq d/2.
 \label{fa:eq-output-implication}
\end{equation}
Define the stopband component of the filter output \(y_n\) by
\begin{equation}
 \widetilde y_n\triangleq \|\mathbf x\|_2\int_{\mathcal B^c}
       p(\e^{-2\pi if})X_N(f)\e^{2\pi inf}\dd f.
 \label{fa:eq-stopband-output}
\end{equation}
Note that
\begin{align}
 |y_n-\widetilde y_n|
 &=\left|\|\mathbf x\|_2\int_{\mathcal B}
       p(\e^{-2\pi if})X_N(f)\e^{2\pi inf}\dd f\right|
 \nonumber\\
 &=\left|\sum_{\ell=0}^{N-1}x_\ell\int_{\mathcal B}
       p(\e^{-2\pi if})\e^{2\pi i(n-\ell)f}\dd f\right|
 \nonumber\\
 &\leq M\sum_{k\in\Z}\left|\int_{\mathcal B}
       p(\e^{-2\pi if})\e^{2\pi ikf}\dd f\right|
 <\frac d4,
 \label{fa:eq-output-comparison}
\end{align}
where the last inequality follows from \eqref{fa:eq-small-filter}. Combining \eqref{fa:eq-output-implication} and \eqref{fa:eq-output-comparison} gives
 \begin{equation}
 w_n\ne0 \Longrightarrow\ \max_{0\leq k<Q}|\widetilde y_{n-k}|>d/4 \Longrightarrow\ \sum_{k=0}^{Q-1}|\widetilde y_{n-k}|^2>\frac{d^2}{16}.
 \end{equation}
Note that \(|w_n|\leq\sum_{k=0}^{Q}|q_k|\,|x_{n-k}|\leq M\). Thus,
\begin{equation}
 |w_n|^2
 \leq\frac{16M^2}{d^2}
       \sum_{k=0}^{Q-1}|\widetilde y_{n-k}|^2,\qquad n\in\Z.
 \label{fa:eq-output-samplewise}
\end{equation}
Summing \eqref{fa:eq-output-samplewise} over \(n\) gives
\begin{equation} 
 \begin{aligned}
 \sum_{n\in\Z}|w_n|^2
 &\leq\frac{16M^2}{d^2}
       \sum_{k=0}^{Q-1}\sum_{n\in\Z}|\widetilde y_{n-k}|^2\\
 &=\frac{16QM^2}{d^2}\sum_{n\in\Z}|\widetilde y_n|^2.
 \end{aligned}
 \label{fa:eq-output-energy-comparison}
\end{equation}
Parseval's theorem gives 
\begin{equation} \label{eq:wyp}
      \begin{aligned}
            \sum_{n\in\Z}|w_n|^2 =&\ \|\mathbf x\|_2^2\int_0^1|q(\e^{-2\pi if})|^2|X_N(f)|^2\dd f,\\
\sum_{n\in\Z}|\widetilde y_n|^2 =&\ \|\mathbf x\|_2^2\int_{\mathcal B^c}|p(\e^{-2\pi if})|^2|X_N(f)|^2\dd f.
      \end{aligned}
 \end{equation}
Combining \eqref{fa:eq-output-energy-comparison} and \eqref{eq:wyp} gives
\begin{equation}
\begin{aligned}
      & \int_0^1|q(\e^{-2\pi if})|^2|X_N(f)|^2\dd f \\
      & \leq\frac{16QM^2}{d^2}
       \int_{\mathcal B^c}|p(\e^{-2\pi if})|^2|X_N(f)|^2\dd f \\
      & \leq\frac{16QM^2}{d^2}
       \sup_{f\in\mathcal B^c}|p(\e^{-2\pi if})|^2L_N,
\end{aligned}
 \label{fa:eq-q-output-parseval}
\end{equation}
which proves \eqref{fa:eq-fixed-filter-control}.
\end{IEEEproof}

We are now ready to prove Theorem~\ref{fa:thm-obstruction}.

\begin{IEEEproof}[Proof of Theorem~\ref{fa:thm-obstruction}]
Take the fixed filter \(q(z)\) and constant \(C\) supplied by
Proposition~\ref{fa:prop-fixed-filter}. Let
\begin{equation}
 \eta\triangleq \frac1\gamma\int_{\mathcal B}
                  |q(\e^{-2\pi if})|^2\dd f.
 \label{fa:eq-positive-target-energy}
\end{equation}
Then \(\eta>0\) because a nonzero FIR filter cannot vanish on \(\mathcal B\). By the Cauchy--Schwarz inequality and the filter bound \eqref{eq:qz_bound}, we obtain
\begin{align}
 &\left|\int_{\mathcal B}|q(\e^{-2\pi if})|^2
       \left(\frac1\gamma-|X_N(f)|^2\right)\dd f\right|
 \nonumber\\
 &\quad\leq
 \left[\int_{\mathcal B}|q(\e^{-2\pi if})|^4\dd f\right]^{1/2}
 \sqrt{\frac{D_N}{\gamma}}
 \nonumber\\
 &\quad\leq
 \left[\int_{\mathcal B}|q(\e^{-2\pi if})|^2\dd f\right]^{1/2}
 \sqrt{\frac{D_N}{\gamma}}
 =\sqrt{\eta D_N}.
 \label{fa:eq-target-output-error}
\end{align}
Then, by the triangle inequality, we have
\begin{equation}
 \begin{aligned}
 \eta-\sqrt{\eta D_N}
 &\leq\int_{\mathcal B}|q(\e^{-2\pi if})|^2|X_N(f)|^2\dd f\\
 &\leq\int_0^1|q(\e^{-2\pi if})|^2|X_N(f)|^2\dd f
          \\
 &\leq CL_N,
 \end{aligned}
 \label{fa:eq-target-output-comparison}
\end{equation}
where the last inequality is \eqref{fa:eq-fixed-filter-control}. Suppose there is a family of nonzero sequences \(\mathbf x^{(N)}\in\mathcal U^N\), \(N=1,2,\ldots\), such that \(L_N\to0\) and \(F_N\to0\). Then \(D_N\to0\) as well by \eqref{m:centering}. Since \(\eta\) and \(C\) are fixed by \(\mathcal U\) and \(\mathcal B\), the left-hand side of \eqref{fa:eq-target-output-comparison} tends to \(\eta>0\) while the right-hand side tends to zero. This contradiction completes the proof of the theorem.
\end{IEEEproof}

\section{Flat-Top Beam Synthesis with GSC Sequences}
\label{m:construction}
In the last section, we ruled out the possibility of approaching ideal flat-top beams with any
fixed finite alphabet. Then a natural question arises: can both the passband ripple and the stopband leakage vanish if the alphabet is infinite? There could be infinitely many choices of amplitudes and phases. This section shows that a unimodular alphabet with infinitely many choices of phases is sufficient to give a positive answer.

\subsection{GSC sequences and the main result}
In our previous work~\cite{DuJiang2024}, we heuristically constructed a family of unimodular sequences with tunable bandwidth.
\begin{definition}
\label{def:gsc}
Let \(m,K\in\Z^+\), \(N=mK\), \(\frac{1}{N}\leq \gamma\leq1\), and \(b\in\mathbb R\).
The generalized step-chirp (GSC) sequence \(\mathbf x=(x_0,\ldots,x_{N-1})\in\mathbb C^N\) is defined by
\begin{equation}
 \begin{aligned}
 x_n&=\exp\left\{\frac{2\pi i\gamma}{K}
       \left[n(k+b)-\frac m2k(k+1)\right]\right\},\\
 n&=km+\ell,\quad k\in\Z_K,\quad
 \ell\in\Z_m,
 \end{aligned}
 \label{m:gsc-original}
\end{equation}
with its passband and stopband specified by
\begin{equation}
 \begin{aligned}
 \mathcal B
 &=\left[\frac{\gamma}{K}(b-\tfrac12),
          \frac{\gamma}{K}(b-\tfrac12)+\gamma\right)\pmod1,\\
 \mathcal B^c&=[0,1)\setminus\mathcal B.
 \end{aligned}
 \label{m:gsc-target-band}
\end{equation}
\end{definition}

Because the parameter $b$ only introduces a shift of the spectrum of $\mathbf x$, we take \(b=1/2\) and
\(\mathcal B=[0,\gamma)\) without loss of generality to investigate the passband ripple $F_N$ and the stopband leakage $L_N$.

Put \(a\triangleq \gamma m/K\), which is the step size of the step-chirp signals defined in~\cite{DuJiang2024}. The sequence to be analyzed is therefore
\begin{equation}
 x_{km+\ell}=\exp\left\{i\pi ak^2+
            \frac{2\pi ia}{m}(k+\tfrac12)\ell\right\}.
 \label{m:sequence}
\end{equation}
For GSC sequences, the passband width $\gamma$ and the step size $a$ may depend on \(N\); we suppress this dependence in the notation.

For a real number \(t\) and a nonempty set \(\mathcal S\subseteq\mathbb R\), define the distance from \(t\) to \(\mathcal S\) by
\[
 \dist(t,\mathcal S)\triangleq\inf_{s\in\mathcal S}|t-s|.
\]
The following theorem characterizes a necessary and sufficient condition for GSC sequences to approach ideal flat-top beams.

\begin{theorem}
\label{thm:vanishing}
For the GSC sequences given in \eqref{m:sequence} with \(a\) bounded, the following statements hold:
\begin{equation}
 \begin{gathered}
 F_N\to0\quad\hbox{and}\quad L_N\to0\\
 \Longleftrightarrow\\
 N\gamma\to\infty,\qquad
 \dist\!\left(a,\{0,1,\tfrac12,\tfrac13,\ldots\}\right)\to0.
 \end{gathered}
 \label{m:classification}
\end{equation}
Furthermore, if \(a\in \{1,\tfrac12,\tfrac13,\ldots\}\) and \(N\gamma\to\infty\), then
\begin{equation}
 F_N=O\!\left((N\gamma)^{-\frac{1}{2}}\right),\quad
 L_N=O\!\left((N\gamma)^{-\frac{1}{2}}\right).
\label{rate:eq-uniform-order}
\end{equation}
\end{theorem}

The remaining part of this section is devoted to the proof of Theorem~\ref{thm:vanishing}.

\subsection{Fourier representations of the discrete and continuous signals}
Define a continuous step-chirp signal $s_a(t)$ by
\begin{equation}
 s_a(t) \triangleq\exp\{i\pi a[k^2+(2k+1)y]\}, \quad t\in \mathbb R,
 \label{m:step}
\end{equation}
where $k=\lfloor t\rfloor$ and $y=t-k$. One may check that $s_a(t)$ is continuous and \(s_a(-t)=s_a(t)\). The GSC sequence in \eqref{m:sequence} is obtained exactly by sampling a truncated version of $s_a(t)$:
\[
 x_{km+\ell}=s_a(k+\ell/m),\qquad 0\leq k<K,\quad 0\leq\ell<m.
\]
The Fourier transform of $s_a(t)$ is defined by
\[
 S_a(f) \triangleq \int_{-\infty}^{\infty} s_a(t) e^{-2\pi i f t} \, dt.
\]

Let \(P\triangleq K/\gamma=m/a\). Next, we adopt the scaling
\(f=a\nu\) for the Fourier transform \(S_a(f)\), and the scaling
\(f=\gamma\nu/K\) for the DTFT \(X_N(f)\) (see \eqref{m:spectrum}).
The latter maps the passband \(f\in[0,\gamma)\) and the
stopband \(f\in[\gamma,1)\) to \(\nu\in[0,K)\) and
\(\nu\in[K,P)\), respectively. The following lemma expresses the rescaled spectra as sums over shifted kernel functions, which will be instrumental in analyzing the asymptotic behavior of the GSC sequences.

\begin{lemma}
\label{m:prop-fourier-representations}
Define two rescaled spectra
\begin{equation}
\begin{aligned}
      \widetilde{X}_N(\nu) \triangleq&\ \sqrt\gamma\,\e^{-i\pi a(1/4-\nu^2)} X_N(\gamma\nu/K), \\
      \widetilde{S}_a(\nu) \triangleq&\ \sqrt a\,\e^{-i\pi a(1/4-\nu^2)} S_a(a\nu).
\end{aligned}
\end{equation}
Then we have
\begin{subequations}
\begin{align}
 \widetilde{X}_N(\nu)&=\sum_{k=0}^{K-1}\Phi_{a,m}(\nu_k),
 \label{m:exact-sum}\\
 \widetilde{S}_a(\nu)&=\sum_{k\in\Z}\Phi_a(\nu_k),
 \label{z:eq-Za}
\end{align}
\end{subequations}
where \(\nu_k=k+\tfrac12-\nu\) and the kernel functions $\Phi_{a,m}$ and $\Phi_a$ are given by
\begin{equation}
 \begin{aligned}
 &\Phi_{a,m}(x)\\[-1mm]
 &\quad=\begin{cases}
 \sqrt a\,\e^{i\pi ax(x-1/m)}
       \dfrac{\sin(\pi ax)}{m\sin(\pi ax/m)},
       &x\notin P\Z,\\[2mm]
 \sqrt a\,\e^{i\pi mj(jP+1)},
       &x=jP,\quad j\in\Z,
 \end{cases}\\[2mm]
 &\Phi_a(x)=\begin{cases}
 \sqrt a\,\e^{i\pi ax^2}\sinc(ax),&x\neq0,\\[1mm]
 \sqrt a,&x=0.
 \end{cases}
 \end{aligned}
 \label{m:finite-kernel}
\end{equation}
\end{lemma}

The proof is direct, thus omitted. It is also straightforward to verify that
\begin{equation} \label{eq:phi_am_period}
 \Phi_{a,m}(\nu_k+P)
 =\e^{i\pi m(P-2\nu)}\Phi_{a,m}(\nu_k).
\end{equation}
\begin{equation}
|\widetilde{X}_N(\nu+P)| = |\widetilde{X}_N(\nu)|,
\end{equation}
\begin{equation}
\widetilde{S}_a(\nu+1) = \widetilde{S}_a(\nu), \quad \widetilde{S}_a(\nu) = \widetilde{S}_a(-\nu)
\end{equation}
and \(\lim_{m\to\infty}\Phi_{a,m}(x)=\Phi_a(x)\) for fixed \(a,x\). Indeed, the kernel function \(\Phi_a(x)\) is the continuous counterpart of \(\Phi_{a,m}(x)\).

\subsection{Stopband and passband estimates}

Define
\begin{equation}
 g_a(x)\triangleq \e^{i\pi ax(x+1)},
 \label{z:eq-ga}
\end{equation}
so that the kernel functions in \eqref{m:finite-kernel} have the following representations for \(x\notin P\Z\) and \(x\neq0\), respectively:
\begin{subequations}
\label{z:eq-kernel-difference}
\begin{align}
       \Phi_{a,m}(x)&=\frac{g_a(x)-g_a(x-1)}{2i\sqrt a}\,
       \frac{\e^{-i\pi x/P}}{P\sin(\pi x/P)}, 
       \label{z:eq-kernel-difference-finite}\\
 \Phi_a(x)&=\frac{g_a(x)-g_a(x-1)}{2\pi i\sqrt a}\,\frac1x.
 \label{z:eq-kernel-difference-continuous}
\end{align}
\end{subequations}

The following two lemmas provide the kernel estimates used extensively in this section, forming the foundation for the subsequent stopband and passband analysis.

\begin{lemma}
\label{z:lem-abel-tail}
Let \(R\triangleq\min_{p\leq k\leq q}|\nu_k|\), where $p, q$ may be infinite.
There exists a constant \(C\) such that
\begin{equation}
 \left|\sum_{k=p}^q\Phi_a(\nu_k)\right| \leq\frac{C}{1+\sqrt a\,R}.
 \label{z:eq-uniform-kernel-block}
\end{equation}
If additionally \(|\nu_k|\leq \frac{P}{2}\) for every \(p\leq k\leq q\), then
\begin{equation}
      \label{eq:kernel_bound_discret}
  \left|\sum_{k=p}^q\Phi_{a,m}(\nu_k)\right| \leq\frac{C}{1+\sqrt a\,R}.
  \end{equation}
\end{lemma}

The proof is given in Appendix~\ref{app:abel-tail}.

\begin{lemma}
\label{z:lem-correction-weight}
For any integers \(p\leq q\) with \(-P<\nu_p\leq \nu_q<P\), there exists a constant \(C\) such that
\begin{equation}
 \begin{aligned}
 &\left|\sum_{k=p}^q\bigl(\Phi_{a,m}(\nu_k)-\Phi_a(\nu_k)\bigr)\right|\\
 &\quad\leq C\left[
 \frac1{1+\sqrt a(P-\nu_q)}+
 \frac1{1+\sqrt a(P+\nu_p)}\right].
 \end{aligned}
 \label{z:eq-uniform-finite-kernel-correction-bound}
\end{equation}
\end{lemma}

The proof is given in Appendix~\ref{app:correction-weight}.

With Lemmas~\ref{z:lem-abel-tail} and~\ref{z:lem-correction-weight} in hand, we now establish Propositions~\ref{z:prop-stopband} and~\ref{z:prop-cellwise} to estimate the magnitude of $\widetilde{X}_N(\nu)$ in the stopband and the difference between $\widetilde{X}_N(\nu)$ and $\widetilde{S}_a(\nu)$ in the passband, respectively.

\begin{samepage}
\begin{proposition}
\label{z:prop-stopband}
There exists a constant \(C\) such that for each \(\nu\in [K,P)\),
\begin{equation}
 |\widetilde{X}_N(\nu)|\leq C\left[
 \frac1{1+\sqrt a(\nu-K)}+
 \frac1{1+\sqrt a(P-\nu)}\right].
 \label{z:eq-stopband-envelope}
\end{equation}
\end{proposition}
\end{samepage}

\begin{IEEEproof}
By \eqref{m:exact-sum} and \eqref{eq:phi_am_period}, we have
\begin{equation} \label{eq:znnu_stopband}
|\widetilde{X}_N(\nu)|
 \leq{}\left|\sum_{\nu_k\geq-\frac{P}{2}}\Phi_{a,m}(\nu_k)\right|
 +\left|\sum_{\nu_k<-\frac{P}{2}}\Phi_{a,m}(\nu_k+P)\right|.
\end{equation}
For each \(\nu\in [K,P)\), we have \(-P< \nu_k <0\) since $\nu_k = k+\tfrac12-\nu$ with \(k\in\Z_K\). If $\nu_k \in (-P,-\frac{P}{2})$, then $ P-\nu< |\nu_k+P| < \frac{P}{2}$. If $\nu_k \in [-\frac{P}{2}, 0)$, then $\nu-K<|\nu_k| \leq \frac{P}{2}$. Applying \eqref{eq:kernel_bound_discret} to \eqref{eq:znnu_stopband} gives
\begin{equation}
     |\widetilde{X}_N(\nu)|
 \leq{}\frac{C}{1+\sqrt a(\nu-K)}
       +\frac{C}{1+\sqrt a(P-\nu)}.
\end{equation}
\end{IEEEproof}

\begin{samepage}
\begin{proposition}
\label{z:prop-cellwise}
There exists a constant \(C\) such that for each \(\nu\in[0,K]\),
\begin{equation}
 \begin{aligned}
 &|\widetilde{X}_N(\nu)-\widetilde{S}_a(\nu)|\\
 &\quad\leq C\left[
 \frac1{1+\sqrt a\,\nu}+\frac1{1+\sqrt a(K-\nu)}
 \right].
 \end{aligned}
 \label{z:eq-passband-envelope}
\end{equation}
\end{proposition}
\end{samepage}

\begin{IEEEproof}
Fix \(\nu\in[0,K]\). Subtracting \eqref{z:eq-Za} from \eqref{m:exact-sum} gives
\begin{equation}
 \widetilde{X}_N(\nu)-\widetilde{S}_a(\nu)
 =\sum_{k=0}^{K-1}\Delta(\nu_k)
 -\sum_{k\in\Z\setminus\Z_K}\Phi_a(\nu_k),
 \label{z:eq-passband-error-decomposition}
\end{equation}
where \(\Delta(x)=\Phi_{a,m}(x)-\Phi_a(x)\).

Let \(C_{\Phi}\) and \(C_d\) denote the constants in
\eqref{z:eq-uniform-kernel-block} and
\eqref{z:eq-uniform-finite-kernel-correction-bound}, respectively. Note that $-P<\nu_0\leq \nu_{K-1}<P$, $P-\nu_{K-1} > \nu$, and $P+\nu_0> K-\nu$.
Hence, by Lemma~\ref{z:lem-correction-weight}, we have
\begin{equation} \label{eq:bound_body}
      \left|\sum_{k=0}^{K-1}\Delta(\nu_k)\right| \leq 
 \frac{C_d}{1+\sqrt a\nu}+
 \frac{C_d}{1+\sqrt a(K-\nu)}.
\end{equation}
Note that $|\nu_k| > \nu$ for $k\leq -1$ and $|\nu_k| > K-\nu$ for $k\geq K$. Hence, by Lemma~\ref{z:lem-abel-tail}, we have
\begin{equation} \label{eq:bound_tails}
      \left|\sum_{k\in\Z\setminus\Z_K}\Phi_a(\nu_k)\right|  \leq 
 \frac{ C_{\Phi}}{1+\sqrt a\nu}+
 \frac{ C_{\Phi}}{1+\sqrt a(K-\nu)}.
\end{equation}
Combining \eqref{z:eq-passband-error-decomposition},
\eqref{eq:bound_body}, and \eqref{eq:bound_tails} proves
\eqref{z:eq-passband-envelope} with \(C=C_d+C_{\Phi}\).
\end{IEEEproof}

\begin{figure}[!htb]
\centering
\includegraphics[width=\columnwidth]{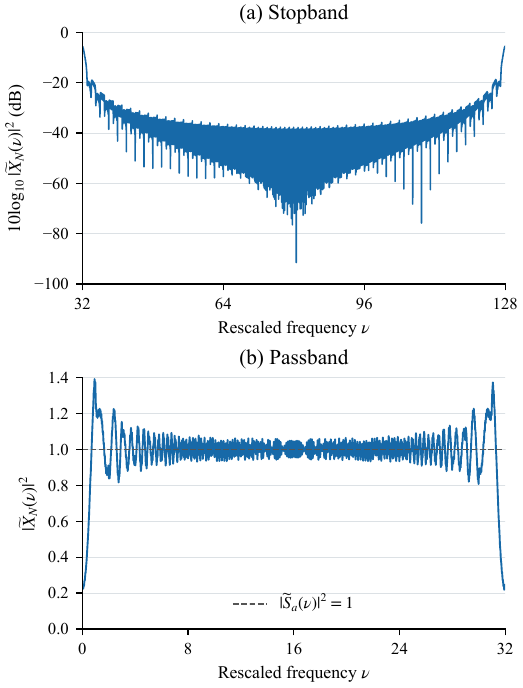}
\caption{GSC power spectrum \(|\widetilde{X}_N(\nu)|^2\) with \(\gamma=1/4\),
\(K=32\), \(m=64\), and \(a=1/2\).
(a) Stopband \([K,P)\), in dB, where \(P=128\).
(b) Passband \([0,K)\), in linear power.}
\label{fig:gsc-spectral-localization}
\end{figure}

\begin{remark}
\label{z:rem-spectral-localization}
Proposition~\ref{z:prop-stopband} predicts the decay of the power spectrum $|\widetilde{X}_N(\nu)|^2$ in the stopband, as shown in
Fig.~\ref{fig:gsc-spectral-localization}(a): the denominators
\(1+\sqrt a(\nu-K)\) and \(1+\sqrt a(P-\nu)\) in
\eqref{z:eq-stopband-envelope} grow with the distances from the band edges $\nu-K$ and $P-\nu$, respectively, consistent with the stopband spectrum
decreasing toward the middle and rising again near the opposite edge.
Proposition~\ref{z:prop-cellwise} implies the convergence of $|\widetilde{X}_N(\nu)|^2$ to $|\widetilde{S}_a(\nu)|^2$ in the interior of the passband as $\sqrt{a}K \to \infty$, as shown in Fig.~\ref{fig:gsc-spectral-localization}(b): the denominators
\(1+\sqrt a\,\nu\) and \(1+\sqrt a(K-\nu)\) in
\eqref{z:eq-passband-envelope} make the bound on
\(|\widetilde{X}_N(\nu)-\widetilde{S}_a(\nu)|\) smallest at the passband center
and larger near either edge, explaining the nearly flat interior
and the stronger deviations near the edges with
\(|\widetilde{S}_{1/2}(\nu)|^2=1\) (which will be proved in the next subsection).
\end{remark}

Remark~\ref{z:rem-spectral-localization} gives a qualitative
interpretation of the spectral estimates in Propositions~\ref{z:prop-stopband} and~\ref{z:prop-cellwise}. The following proposition further
quantifies this behavior.
\begin{proposition}
\label{z:prop-uniform-comparison}
For \(N\gamma\) sufficiently large, there exist constants \(C_1,C_2\) such that
\begin{align}
 L_N&\leq C_1(N\gamma)^{-1/2},
 \label{z:eq-leakage-rate}\\
 |F_N-V(a)|
 &\leq C_2\left[
 \sqrt{V(a)}(N\gamma)^{-1/4}
 +(N\gamma)^{-1/2}\right],
 \label{z:eq-ripple-rate}
\end{align}
where 
\begin{equation}
 V(a)=\int_0^1\bigl(|\widetilde{S}_a(\nu)|^2-1\bigr)^2\dd \nu.
 \label{z:eq-V}
\end{equation}
Consequently,
\begin{equation}
 N\gamma\to\infty\quad\Longrightarrow\quad
 L_N\to0,\qquad F_N\to V(a).
 \label{z:eq-common-zero}
\end{equation}
\end{proposition}

\begin{IEEEproof}
By \(|\widetilde{X}_N(\nu)|^2=\gamma|X_N(\gamma\nu/K)|^2\), \eqref{m:leakage} and \eqref{m:ripple} can be rewritten as
\begin{subequations}
       \begin{align}
 L_N&=\frac1K\int_K^P|\widetilde{X}_N(\nu)|^2\dd\nu,\label{eq:LZ} \\ 
 F_N&=\frac1K\int_0^K
 \left(\frac{|\widetilde{X}_N(\nu)|^2}{1-L_N}-1\right)^2\dd\nu. \label{eq:FZ}
 \end{align}
\end{subequations}

Choose a common constant \(C_0\) for
Lemma~\ref{z:lem-abel-tail} and
Propositions~\ref{z:prop-stopband} and \ref{z:prop-cellwise}.
Applying Proposition~\ref{z:prop-stopband} to \eqref{eq:LZ} yields
\[
 \begin{aligned}
 L_N&\leq\frac{2C_0^2}{K}\int_K^P
 \frac{\dd\nu}{(1+\sqrt a(\nu-K))^2}\\
 &\quad+\frac{2C_0^2}{K}\int_K^P
 \frac{\dd\nu}{(1+\sqrt a(P-\nu))^2}.
 \end{aligned}
\]
Letting \(t=\nu-K\), we obtain
\begin{equation}
 L_N\leq\frac{4C_0^2}{K}\int_0^{P-K}
       \frac{\dd t}{(1+\sqrt a\,t)^2} \leq\frac{4C_0^2}{K\sqrt a}=\frac{4C_0^2}{\sqrt{N\gamma}},
 \label{z:eq-leakage-integral}
\end{equation}
which proves \eqref{z:eq-leakage-rate} with $C_1=4C_0^2$.

Define the normalized norm
\[
 \|g\|_K\triangleq\left(\frac1K\int_0^K|g(\nu)|^2\dd\nu\right)^{1/2}.
\]
Similarly, applying Proposition~\ref{z:prop-cellwise} gives
\begin{equation}
 \|\widetilde{X}_N-\widetilde{S}_a\|_K^2
 \leq\frac{4C_0^2}{K}\int_0^K\frac{\dd\nu}{(1+\sqrt a\,\nu)^2}
 \leq\frac{4C_0^2}{\sqrt{N\gamma}}.
 \label{z:eq-mean-spectral-error}
\end{equation}
Lemma~\ref{z:lem-abel-tail} applied to \eqref{z:eq-Za}, together with
Proposition~\ref{z:prop-cellwise}, gives
\begin{equation} \label{eq:za_za_plus_zn}
 |\widetilde{S}_a(\nu)|\leq C_0,\qquad
 |\widetilde{X}_N(\nu)|+|\widetilde{S}_a(\nu)|\leq4C_0
\end{equation}
on \([0,K]\). Then
\begin{equation} \label{eq:zn2_za2}
 \bigl||\widetilde{X}_N|^2-|\widetilde{S}_a|^2\bigr|
 \leq4C_0|\widetilde{X}_N-\widetilde{S}_a|.
\end{equation}
By the triangle inequality in the sense of \(\|\cdot\|_K\), we have
\begin{equation}
 \begin{aligned}
 &\left\|\frac{|\widetilde{X}_N|^2}{1-L_N}-|\widetilde{S}_a|^2\right\|_K\\
\leq&\ \frac{\bigl\||\widetilde{X}_N|^2-|\widetilde{S}_a|^2\bigr\|_K
                   +L_N\bigl\||\widetilde{S}_a|^2\bigr\|_K}{|1-L_N|}\\
 \leq&\ \frac{4C_0\|\widetilde{X}_N-\widetilde{S}_a\|_K+C_0^2L_N}{|1-L_N|}\\
\leq&\ \frac{8C_0^2(N\gamma)^{-\frac14}+4C_0^4(N\gamma)^{-\frac12}}{|1-L_N|}\\
\leq&\ \frac{(8C_0^2+4C_0^4)(N\gamma)^{-\frac14}}{|1-L_N|},
 \end{aligned}
 \label{z:eq-normalized-power-error}
\end{equation}
where the second inequality follows from \eqref{eq:zn2_za2} and \eqref{eq:za_za_plus_zn}, and the third inequality follows from \eqref{z:eq-mean-spectral-error} and \eqref{z:eq-leakage-integral}.
By \eqref{z:eq-leakage-integral}, if $\sqrt{N\gamma}\geq8C_0^2$, then \(L_N\leq1/2\), and hence
\begin{equation} \label{eq:diff_norm}
      \left\|\frac{|\widetilde{X}_N|^2}{1-L_N}-|\widetilde{S}_a|^2\right\|_K \leq C_3(N\gamma)^{-1/4},
\end{equation}
where \(C_3=16C_0^2+8C_0^4\). Note that
\[
 \sqrt{F_N}=\left\|\frac{|\widetilde{X}_N|^2}{1-L_N}-1\right\|_K,\quad
 \sqrt{V(a)}=\bigl\||\widetilde{S}_a|^2-1\bigr\|_K.
\]
The reverse triangle inequality therefore gives
\[
 \bigl|\sqrt{F_N}-\sqrt{V(a)}\bigr|\leq \left\|\frac{|\widetilde{X}_N|^2}{1-L_N}-|\widetilde{S}_a|^2\right\|_K\leq C_3(N\gamma)^{-1/4}.
\]
Consequently,
\[
|F_N-V(a)|\leq 2C_3\sqrt{V(a)}\,(N\gamma)^{-1/4}
          +C_3^2(N\gamma)^{-1/2}.
\]
Taking \(C_2=\max\{2C_3,C_3^2\}\) proves \eqref{z:eq-ripple-rate}. By \eqref{z:eq-V} and \eqref{eq:za_za_plus_zn}, we have $V(a)\leq(C_0^2+1)^2$. The boundedness of \(V(a)\) together with \eqref{z:eq-leakage-rate} and \eqref{z:eq-ripple-rate} then gives
\eqref{z:eq-common-zero}.
\end{IEEEproof}

By \eqref{z:eq-common-zero}, $\lim_{N\gamma\to\infty} (F_N+L_N)=0$ is equivalent to \(V(a)\to0\). Next, we give the zero set and establish the continuity of \(V(a)\).


\subsection{\(V(a)\)'s zero set and continuity}
Define the Fourier coefficient of $|\widetilde{S}_a(\nu)|^2$ as
\begin{equation}
 c_h(a)\triangleq\int_0^1|\widetilde{S}_a(\nu)|^2\e^{2\pi ih\nu}\dd \nu,\quad h\in\Z.
\label{z:eq-ch-def}
\end{equation}
Applying Parseval's theorem to
\(|\widetilde{S}_a|^2-1\) in \eqref{z:eq-V} gives
\begin{equation}
 V(a)=|c_0(a)-1|^2+2\sum_{h=1}^{\infty}|c_h(a)|^2.
 \label{z:eq-V-series}
\end{equation}
To determine the zero set of $V(a)$, we next compute \(c_h(a)\) for \(h\geq0\).
\begin{lemma}
\label{z:prop-fourier-coefficients}
For \(h\geq0\) with \(\kappa_h\triangleq\lfloor h/a\rfloor\),
\begin{equation}
 c_h(a)=(-1)^h
 \begin{cases}
  1-h/a,&\kappa_h=0,\\[1mm]
  \sinc\!\bigl(a\kappa_h(\kappa_h+1)\bigr),&\kappa_h\geq1.
 \end{cases}
 \label{z:eq-ch-formula}
\end{equation}
\end{lemma}

The proof is given in Appendix~\ref{app:fourier-coefficients}.



\begin{proposition}
\label{z:coro-V}
\begin{equation}
 V(a)=0
 \quad\Longleftrightarrow\quad
 a\in\left\{1,\frac12,\frac13,\ldots\right\}.
 \label{z:eq-V-zero-set}
\end{equation}
\end{proposition}

\begin{IEEEproof}
By Lemma~\ref{z:prop-fourier-coefficients}, $c_0(a)=1$. Hence, by \eqref{z:eq-V-series},
\begin{equation} \label{eq:v_to_c}
       V(a) =0 \quad\Longleftrightarrow\quad c_h(a)=0\ , \forall h\geq1.
\end{equation}

If \(a=1/J\) with \(J\in\Z^+\), then $\kappa_h=Jh$. By Lemma~\ref{z:prop-fourier-coefficients}, \(c_h(a)=0\) for every \(h\geq1\). Conversely, if \(a>1\), then \(c_1(a)=1/a-1\ne0\); if \(0<a<1\) is not an integer reciprocal, then for $\kappa_1 = \lfloor 1/a \rfloor$ we have $1\leq \kappa_1<a\kappa_1(\kappa_1+1)<\kappa_1+1$, so \(c_1(a)=-\sinc(a\kappa_1(\kappa_1+1))\ne0\). Therefore,
\begin{equation} \label{eq:c_to_a}
      c_h(a)=0\ , \forall h\geq1 \quad\Longleftrightarrow\quad a\in\left\{1,\frac12,\frac13,\ldots\right\}.
\end{equation}
Combining \eqref{eq:v_to_c} and \eqref{eq:c_to_a} completes the proof.
\end{IEEEproof}

\begin{proposition}
\label{z:prop-V-continuity}
The function \(V(a)\) is continuous on \((0,\infty)\) and satisfies
\(
 \lim_{a\to0^+}V(a)=0.
\)
\end{proposition}

\begin{IEEEproof}
By Lemma~\ref{z:prop-fourier-coefficients}, \(c_0(a)=1\) and is thus continuous.
For \(h\in\Z^+\), $c_h(a)$ is also continuous: at any possible discontinuity point \(a=h/j\), \(j\in\Z^+\), the one-sided limits are
\[
 \begin{aligned}
 \lim_{a\to(h/j)^-}c_h(a)
 &=(-1)^h\sinc\!\bigl(h(j+1)\bigr)=0,\\
 \lim_{a\to(h/j)^+}c_h(a)
 &=\begin{cases}
 (-1)^h\sinc\!\bigl(h(j-1)\bigr)=0,&j\geq2,\\
 (-1)^h(1-h/h)=0,&j=1.
 \end{cases}
 \end{aligned}
\]

For \(h\geq2a\), \(\kappa_h=\lfloor h/a\rfloor  \geq h/a-1 \geq h/(2a)\) and \(\kappa_h+1\geq h/a\), so
\[
 |c_h(a)|\leq\frac1{\pi a\kappa_h(\kappa_h+1)}\leq\frac{2a}{\pi h^2}.
\]
If $a\in (0, A]$, then $|c_h(a)|^2\leq \frac{4A^2}{\pi^2 h^4}$ for $h\geq 2A$. Hence the series in
\eqref{z:eq-V-series} converges uniformly on $(0, A]$ by the Weierstrass \(M\)-test. Together with the continuity of $c_h(a)$, this implies the continuity of \(V(a)\) on \((0,A]\), and thus on \((0,\infty)\) since \(A>0\) was arbitrary. Moreover, for \(0<a\leq1/2\),
\[
 0\leq V(a)\leq\frac{8a^2}{\pi^2}\sum_{h=1}^{\infty}\frac1{h^4}
 =\frac{4\pi^2}{45}a^2,
\]
thus \(\lim_{a\to0^+}V(a)=0\).
\end{IEEEproof}

\subsection{A necessary condition for asymptotically vanishing leakage}
\label{z:sec-classification}
The remaining ingredient for proving Theorem~\ref{thm:vanishing} is the following necessary condition.
\begin{proposition}
\label{z:prop-dpss}
If \(L_N\to0\), then \(N\gamma\to\infty\).
\end{proposition}

\begin{IEEEproof}
By the Fourier correspondence
\[
 \int_0^1|X_N(f)|^2\e^{2\pi ikf}\dd f
 =\frac{r_k}N, \quad k=0,\ldots,N-1
\]
we have 
\begin{equation} \label{eq:p_bound}
       \begin{aligned}
 \frac1N\Re\!\left(r_k\right) =&\ \int_0^1\cos(2\pi kf)|X_N(f)|^2\dd f\\
 \geq&\ \left(1-2\pi^2k^2\gamma^2\right)(1-L_N)-L_N\\
\geq&\ 1-2\pi^2k^2\gamma^2-2L_N.
 \end{aligned}
\end{equation}
where the first inequality uses
\(\cos t\geq1-t^2/2\) on the passband $f\in[0,\gamma]$ and \(\cos t\geq-1\)
on the stopband $f\in[\gamma,1]$. On the other hand, we have $\Re(r_k)\leq|r_k|\leq N-k$. Thus,
\begin{equation}
 L_N \geq \frac k{2N} -\pi^2k^2\gamma^2.
 \label{z:eq-autocorrelation-leakage}
\end{equation}

If $N\gamma\not\to\infty$, then along a subsequence $N=N_j\to\infty$ we have $N\gamma\leq\Lambda$ for some fixed \(\Lambda\geq1\).
Take \(k=\lfloor N/(4\pi^2\Lambda^2)\rfloor<N\). Since
\(0\leq k/N\leq1/(4\pi^2\Lambda^2)\), we obtain
\begin{equation}
 L_N\geq\frac kN\left(\frac12-\pi^2\Lambda^2\frac kN\right)
    \geq\frac{k}{4N}.
\end{equation}
Moreover, \(k/N\to1/(4\pi^2\Lambda^2)\) along this subsequence. Thus,
\[
 \liminf_{j\to\infty}L_{N_j}\geq\frac1{16\pi^2\Lambda^2}>0,
\]
contradicting \(L_N\to0\). Hence \(N\gamma\to\infty\).
\end{IEEEproof}

We are now ready to prove Theorem~\ref{thm:vanishing}.
\begin{IEEEproof}[Proof of Theorem~\ref{thm:vanishing}]
First, Proposition~\ref{z:prop-dpss} together with \eqref{z:eq-common-zero} in Proposition~\ref{z:prop-uniform-comparison} shows that 
\begin{equation}
 F_N\to0,\; L_N\to0\;\Longleftrightarrow\; N\gamma\to\infty,\; V(a)\to0.
 \label{z:eq-error-reduction}
\end{equation}
Second, Proposition~\ref{z:coro-V} and Proposition~\ref{z:prop-V-continuity} imply
\begin{equation}
 V(a)\to0
 \ \Longleftrightarrow\ 
 \dist\!\left(a,\{0,1,\tfrac12,\tfrac13,\ldots\}\right)\to0.
 \label{z:eq-V-distance}
\end{equation}
Combining \eqref{z:eq-error-reduction} and \eqref{z:eq-V-distance}
proves \eqref{m:classification}.

Finally, if \(a\in\{1,\tfrac12,\tfrac13,\ldots\}\), then
\(V(a)=0\) by Proposition~\ref{z:coro-V}.
Substituting this into \eqref{z:eq-leakage-rate}--\eqref{z:eq-ripple-rate} proves \eqref{rate:eq-uniform-order}.
\end{IEEEproof}

\subsection{Numerical Results}
\label{m:numerical-results}
\begin{figure}[!htbp]
 \centering
 \includegraphics[width=\columnwidth]{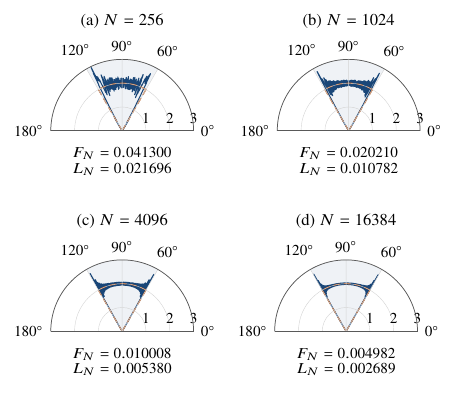}
 \caption{GSC beam convergence with \(a=1/2\), \(\gamma=1/2\), and \(m=K\). Panels (a)--(d) use \(K\in\{16,32,64,128\}\),
 giving \(N\in\{256,1024,4096,16384\}\), respectively.}
 \label{fig:gsc-beam-convergence}
\end{figure}

\begin{figure}[!htbp]
 \centering
 \includegraphics[width=\columnwidth]{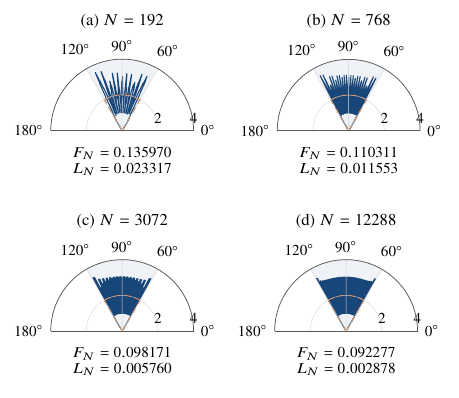}
 \caption{Persistent GSC passband ripple with \(a=2/3\), \(\gamma=1/2\), and \(m=4K/3\). Panels (a)--(d) use
 \(K\in\{12,24,48,96\}\),
 giving \(N\in\{192,768,3072,12288\}\), respectively.}
 \label{fig:gsc-ripple-counterexample}
\end{figure}

To verify Theorem~\ref{thm:vanishing}, Figures~\ref{fig:gsc-beam-convergence} and~\ref{fig:gsc-ripple-counterexample}
compare the asymptotic behavior of GSC beams with $a=\frac{1}{2}$ and $a=\frac{2}{3}$, respectively.
Here the radial coordinate is \(|X_N(\cos\theta/2)|^2\). The blue curve is the GSC beam, while the orange dashed curve is
the ideal flat-top beam with gain two. Shading marks the target sector
\(60^\circ\leq\theta\leq120^\circ\). We set \(\gamma=1/2\) and \(b=(1-K)/2\) to center the passband at \([-1/4,1/4)\pmod1\). The reported \(F_N\) and \(L_N\) are computed from the
autocorrelation closed forms in
Proposition~\ref{m:prop-correlation-criteria}.

In Figure~\ref{fig:gsc-beam-convergence}, the beams converge as \(N\) increases because $a=\frac{1}{2}$ is an integer reciprocal. Specifically, quadrupling \(N\gamma\) approximately halves both
\(F_N\) and \(L_N\), consistent with the proved
upper bound of \(O((N\gamma)^{-1/2})\). In contrast, in Figure~\ref{fig:gsc-ripple-counterexample}, quadrupling \(N\gamma\) approximately halves the stopband leakage \(L_N\), while the passband ripple $F_N$ persists as $N$ increases, because $\dist\!\left(\frac{2}{3},\{0,1,\tfrac12,\tfrac13,\ldots\}\right)=\frac{1}{6}$, violating the condition $\dist\!\left(a,\{0,1,\tfrac12,\tfrac13,\ldots\}\right)\to0$.

\FloatBarrier

\section{Conclusion and Future Directions}
 To answer the fundamental question about the feasibility of approaching ideal flat-top beams, this work broadened the scope of classical sequence design beyond high merit-factor sequences. In particular, we proved that no sequence with a finite alphabet permits the passband ripple and stopband leakage to tend to zero simultaneously. We also proved that the passband ripple and stopband leakage of the GSC sequences, among unimodular sequences with infinite alphabets, could tend to zero simultaneously under some necessary and sufficient conditions, and established the rate of convergence. 

We hope this work will open new avenues for research in sequence design. Future work may include the following two directions:
\begin{enumerate}
 \item 
 Determine whether unimodular sequences can achieve simultaneous
 decay of \(F_N\) and \(L_N\) faster than \((N\gamma)^{-1/2}\)
 as \(N\gamma\to\infty\).

 \item 
 For a polyphase alphabet, such as the binary or
 quaternary alphabet, derive lower bounds on \(F_N\) under a
 prescribed leakage constraint \(L_N\leq\epsilon_2\), and determine
 whether these bounds can be attained or approached by explicit constructions of sequences over the alphabet.
\end{enumerate}

\appendices
\section{Proof of Lemma~\ref{fa:lem-stable-filter}}
\label{app:finite-alphabet}

\begin{IEEEproof}
Write the closure of \(\mathcal B\) as \([\alpha,\alpha+\gamma]\) modulo one, and set
\[
 I\triangleq\{\e^{-2\pi if}:\alpha\leq f\leq\alpha+\gamma\}.
\]
Vaidyanathan's FIR prediction-error construction
\cite[Sec.~II, eq.~(2)]{Vaidyanathan1987} provides
one fixed polynomial \(h(z)\) such that
\begin{equation}
 h(0)=1,\qquad
 0<\rho\triangleq\max_{z\in I}|h(z)|<1.
 \label{fa:eq-base-predictor}
\end{equation}
For an integer \(\ell\geq1\), to be chosen below, define
\begin{equation}
 p(z)=(1-\e^{2\pi i\alpha}z)
       (1-\e^{2\pi i(\alpha+\gamma)}z)h(z)^\ell.
 \label{fa:eq-endpoint-zero-predictor}
\end{equation}
Then \(p(0)=1\), 
\begin{equation} \label{eq:end_points}
p(\e^{-2\pi i\alpha})=p(\e^{-2\pi i(\alpha+\gamma)})=0
\end{equation}
and 
\begin{equation}
      \max_{z\in I}|p(z)| \leq4\rho^\ell.
\end{equation}
By the product rule and the triangle inequality,
\begin{equation}
 \max_{z\in I}|p'(z)|\leq4\rho^\ell
       +4\ell\rho^{\ell-1}\max_{z\in I}|h'(z)|,
 \label{fa:eq-filter-derivative-bounds}
\end{equation}
where primes denote derivatives with respect to \(z\).

By \eqref{eq:end_points}, one may check that the function \(p(\e^{-2\pi if})\ind_{\mathcal B}(f)\) is absolutely
continuous and \(\frac{\dd}{\dd f}[p(\e^{-2\pi if})\ind_{\mathcal B}(f)]\in L^2([0,1])\). Thus, we can apply the standard Fourier coefficient estimate~\cite[p.~32, eq.~(6.2)]{Katznelson2004} to obtain
\begin{align}
 &\sum_{k\in\Z}
  \left|\int_{\mathcal B}p(\e^{-2\pi if})\e^{2\pi ikf}\dd f\right|
 \nonumber\\
 &\leq\int_{\mathcal B}|p(\e^{-2\pi if})|\dd f
       +\frac1{\sqrt{12}}
       \left(\int_{\mathcal B}
       \left|\frac{\dd}{\dd f}p(\e^{-2\pi if})\right|^2\dd f\right)^{1/2}
 \nonumber\\
 &\leq\gamma\max_{z\in I}|p(z)|
       +\frac{2\pi\sqrt\gamma}{\sqrt{12}}\max_{z\in I}|p'(z)|
 \nonumber\\
 &\leq4\gamma\rho^\ell
       +\frac{8\pi\sqrt\gamma}{\sqrt{12}}
        \Bigl(\rho^\ell+\ell\rho^{\ell-1}
        \max_{z\in I}|h'(z)|\Bigr)
 \label{fa:eq-filter-l1-bound}
\end{align}
Given \(\varepsilon>0\), since \(0<\rho<1\) and \(\max_{z\in I}|h'(z)|\) is independent of \(\ell\), we can choose a finite \(\ell\) to make the final right-hand side of \eqref{fa:eq-filter-l1-bound} smaller than \(\varepsilon\). This completes the proof.
\end{IEEEproof}

\section{Proof of Lemma~\ref{z:lem-abel-tail}}
\label{app:abel-tail}

\begin{IEEEproof}
Before proving \eqref{z:eq-uniform-kernel-block} and
\eqref{eq:kernel_bound_discret}, assume without loss of
generality that \(\nu_p,\ldots,\nu_q\) are all nonnegative or all
nonpositive. Otherwise, we can always split the sums in \eqref{z:eq-uniform-kernel-block} or
\eqref{eq:kernel_bound_discret} into two parts according to \(\nu_k<0\) and
\(\nu_k\geq0\), and bound their magnitudes separately.

If \(R>0\), then \(\nu_k\neq0\) for \(p\leq k\leq q\), and
\eqref{z:eq-kernel-difference-continuous} gives
\begin{equation}
 \begin{aligned}
 &\left|\sum_{k=p}^q\Phi_a(\nu_k)\right| = \ \frac{1}{2\pi\sqrt a}\left|
       \sum_{k=p}^q\frac{g_a(\nu_k)-g_a(\nu_{k-1})}{\nu_k}\right|\\
 =&\ \frac{1}{2\pi\sqrt a}\Biggl|
       \sum_{k=p}^{q-1}g_a(\nu_k)
       \left(\frac{1}{\nu_k}-\frac{1}{\nu_{k+1}}\right) +\frac{g_a(\nu_q)}{\nu_q}
       -\frac{g_a(\nu_{p-1})}{\nu_p}\Biggr|\\
\leq&\ \frac{1}{2\pi\sqrt a}\left[\sum_{k=p}^{q-1}\left|\frac{1}{\nu_k}-\frac{1}{\nu_{k+1}}\right|
       +\left|\frac{1}{\nu_q}\right|+\left|\frac{1}{\nu_p}\right|\right]= \frac{1}{\pi\sqrt aR}.
 \end{aligned}
 \label{z:eq-abel-tail}
\end{equation}

For $\Phi_{a,m}(x)$ with \(x\notin P\Z\), define
\begin{equation}
 \omega(x) \triangleq \frac{\e^{-i\pi x/P}}{P\sin(\pi x/P)}, \quad \omega^{\prime}(x) \triangleq \frac{d\omega}{dx}.
 \end{equation}
Since \(|\sin(\pi x/P)|\geq2|x|/P\) for \(0<|x|\leq \frac{P}{2}\), we have
\begin{equation}
 \begin{aligned}
 \left|\omega(x)\right|
     &\leq\frac1{2|x|},\\
 \left|\omega^{\prime}(x)\right|
     &=\frac\pi{P^2\sin^2(\pi x/P)}
       \leq\frac\pi{4x^2}.
 \end{aligned}
 \label{z:eq-central-weight-bounds}
\end{equation}
For the kernel $\Phi_{a, m}$, the additional condition
\(|\nu_k|\leq \frac{P}{2}\), together with \(R>0\), ensures that
\([\nu_p,\nu_q]\) contains no point of \(P\Z\). As in \eqref{z:eq-abel-tail}, we have
\begin{equation}
 \begin{aligned}
 &\left|\sum_{k=p}^q\Phi_{a,m}(\nu_k)\right|\\
 \leq&\ \frac{1}{2\sqrt a}\Biggl[
       \sum_{k=p}^{q-1}\left|\omega(\nu_k)-\omega(\nu_{k+1})\right| +\left|\omega(\nu_q)\right|
       +\left|\omega(\nu_p)\right|\Biggr]\\
\leq &\ \frac{1}{2\sqrt a}\left[
       \int_{\nu_p}^{\nu_q}|\omega'(x)|\dd x+\frac1R\right]\\
 \leq &\ \frac{1}{2\sqrt a}\left[
       \frac\pi4\int_R^\infty\frac{\dd x}{x^2}+\frac1R\right]=\ \frac{\pi+4}{8\sqrt a\,R}.
 \end{aligned}
 \label{z:eq-abel-tail-m}
\end{equation}

Next, we prove the bounds \eqref{z:eq-uniform-kernel-block} and
\eqref{eq:kernel_bound_discret}.
Write \(\Psi\) for either \(\Phi_a\) or \(\Phi_{a,m}\).
If \(\sqrt aR\geq1\), by
\eqref{z:eq-abel-tail} and
\eqref{z:eq-abel-tail-m}, there exists a constant \(C\) such that
\[
 \left|\sum_{k=p}^q\Psi(\nu_k)\right|
 \leq\frac{C}{2\sqrt a\,R}
 \leq\frac{C}{1+\sqrt a\,R}.
\]
If \(\sqrt aR<1\), including \(R=0\), define
\[
 \begin{aligned}
 \mathcal I_0&\triangleq\{k\in\{p,\ldots,q\}:|\nu_k|<1/\sqrt a\},\\
 \mathcal I_1&\triangleq\{p,\ldots,q\}\setminus\mathcal I_0,\\
 S_j&\triangleq\sum_{k\in\mathcal I_j}\Psi(\nu_k),\qquad j=0,1.
 \end{aligned}
\]
Note that 
\(\lvert\mathcal I_0\rvert\leq1/\sqrt a+1\) and 
\(|\Psi(x)|\leq\sqrt a\) by \eqref{m:finite-kernel}. Thus,
 \begin{equation}
 |S_0|\leq\sqrt a\,\lvert\mathcal I_0\rvert\leq1+\sqrt a.
 \end{equation}
If \(\mathcal I_1\neq\varnothing\), it can be written as $\{p^{\prime}, p^{\prime}+1, \cdots, q^{\prime}\}$ with $p^{\prime}\geq p$, $q^{\prime}\leq q$ and 
\[
 R_1\triangleq\min_{k\in\mathcal I_1}|\nu_k|\geq1/\sqrt a.
\]
Hence \eqref{z:eq-abel-tail} and \eqref{z:eq-abel-tail-m} yield
\[
 |S_1|\leq
 \begin{cases}
 \dfrac{1}{\pi\sqrt a\,R_1}\leq\dfrac1\pi,
     &\Psi=\Phi_a,\\[2mm]
 \dfrac{\pi+4}{8\sqrt a\,R_1}\leq\dfrac{\pi+4}{8},
     &\Psi=\Phi_{a,m}.
 \end{cases}
\]
For \(\mathcal I_1=\varnothing\), \(S_1=0\). Hence, 
\begin{equation}
      \left|\sum_{k=p}^q\Psi(\nu_k)\right|
 \leq|S_0|+|S_1| \leq1+\sqrt a+\frac{\pi+4}{8}.
\end{equation}
Taking \(C\geq2(1+\sqrt A+(\pi+4)/8)\), where \(A\) is an upper bound for \(a\), we obtain
\[
\left|\sum_{k=p}^q\Psi(\nu_k)\right| \leq\frac C2\leq\frac{C}{1+\sqrt aR}.
\tag*{\IEEEQEDhere}
\]
\end{IEEEproof}

\section{Proof of Lemma~\ref{z:lem-correction-weight}}
\label{app:correction-weight}

\begin{IEEEproof}
For \(\nu_k\neq0\), \eqref{z:eq-kernel-difference} gives
\[
 \begin{aligned}
 &\Phi_{a,m}(\nu_k)-\Phi_a(\nu_k)\\
=&\ \frac{g_a(\nu_k)-g_a(\nu_{k-1})}{2i\sqrt a}
 \left[\frac{\cot(\pi\nu_k/P)-i}{P}
       -\frac1{\pi \nu_k}\right].
 \end{aligned}
\]
To separate the poles of $\cot(\pi x/P)$ at \(\pm P\), define 
\[
 \rho_P(x)\triangleq\frac{\cot(\pi x/P)-i}{P}
       -\sum_{j=-1}^{1}\frac1{\pi(x-jP)}, \quad 0<|x|<P.
\]
Then the difference becomes
\begin{equation}
 \begin{aligned}
 \Phi_{a,m}(\nu_k)-\Phi_a(\nu_k) = &\ \frac{g_a(\nu_k)-g_a(\nu_{k-1})}{2\pi i\sqrt a}
       \sum_{\sigma=\pm1}\frac1{\nu_k+\sigma P}\\
 & +\frac{g_a(\nu_k)-g_a(\nu_{k-1})}{2i\sqrt a}\,\rho_P(\nu_k).
 \end{aligned}
 \label{z:eq-kernel-difference-comparison}
\end{equation}

We first bound the part containing the poles. By \eqref{z:eq-ga} and \eqref{z:eq-kernel-difference}, we have
\begin{equation} \label{eq:g_and_phi}
      \frac{g_a(\nu_k)-g_a(\nu_{k-1})}{2\pi i\sqrt a\,(\nu_k+\sigma P)}
 =\e^{-i\pi mP+2\pi i\sigma m\nu}\Phi_a(\nu_k+\sigma P).
\end{equation}
Then
\begin{equation}
 \begin{aligned}
 &\left|\sum_{k=p}^q
 \frac{g_a(\nu_k)-g_a(\nu_{k-1})}{2\pi i\sqrt a}
 \sum_{\sigma=\pm1}\frac1{\nu_k+\sigma P}\right|\\
 &\quad\leq\sum_{\sigma=\pm1}
       \left|\sum_{k=p}^q\Phi_a(\nu_k+\sigma P)\right|\\
 &\quad\leq C_{\Phi}\left[
 \frac1{1+\sqrt a(P-\nu_q)}+
 \frac1{1+\sqrt a(P+\nu_p)}\right],
 \end{aligned}
 \label{z:eq-correction-pole-bound}
\end{equation}
where the two inequalities follow from \eqref{eq:g_and_phi}  and \eqref{z:eq-uniform-kernel-block}, respectively.

For the remainder in \eqref{z:eq-kernel-difference-comparison}, define
\[
 \begin{aligned}
 \rho_1(t)&\triangleq\cot(\pi t)-\sum_{j=-1}^{1}\frac1{\pi(t-j)}-i,
 \quad 0<|t|<1,\\
 \rho_1(0)&=-i,\qquad \rho_1(\pm1)=\mp\frac3{2\pi}-i.
 \end{aligned}
\]
Using the expansion \(\cot t = t^{-1} -t/3 +O(t^3)\)
and the periodicity \(\cot (t+\pi) = \cot t\), one may check that \(\rho_1\in C^1([-1,1])\).
Thus, the absolute constant
\[
 C_{\rho}\triangleq\sup_{|t|\leq1}|\rho_1(t)|
       +\sup_{|t|\leq1}|\rho_1'(t)|
\]
is finite. The scaling \(\rho_P(x)=P^{-1}\rho_1(x/P)\) extends
\(\rho_P\) to \([-P,P]\), and
\(\rho_P'(x)=P^{-2}\rho_1'(x/P)\) gives
\begin{equation}
 \sup_{|x|\leq P}|\rho_P(x)|
 +P\sup_{|x|\leq P}|\rho_P'(x)|\leq\frac{C_{\rho}}{P}.
 \label{z:eq-correction-remainder-bounds}
\end{equation}
The identity \eqref{z:eq-kernel-difference-comparison} now also holds
at \(\nu_k=0\) by the continuity of $\rho_1(t)$. Then we have
\begin{equation}
 \begin{aligned}
 &\left|\sum_{k=p}^q
 \frac{g_a(\nu_k)-g_a(\nu_{k-1})}{2i\sqrt a}\,\rho_P(\nu_k)\right|\\
 &\quad\leq\frac{|\rho_P(\nu_p)|+|\rho_P(\nu_q)|
       +\int_{\nu_p}^{\nu_q}|\rho_P'(x)|\dd x}{2\sqrt a}
 \leq\frac{C_{\rho}}{\sqrt aP}.
 \end{aligned}
 \label{z:eq-tail-Abel}
\end{equation}

Combining \eqref{z:eq-kernel-difference-comparison},
\eqref{z:eq-correction-pole-bound}, and \eqref{z:eq-tail-Abel}
proves \eqref{z:eq-uniform-finite-kernel-correction-bound}
with \(C=C_{\Phi}+C_{\rho}\), since \(\nu_p\leq \nu_q\) and
\(\sqrt aP\geq1\) imply
\[
 \begin{aligned}
 &\frac1{1+\sqrt a(P-\nu_q)}+
 \frac1{1+\sqrt a(P+\nu_p)}\\
 &\quad\geq\frac2{1+\sqrt aP}\geq\frac1{\sqrt aP}.
 \end{aligned}
 \tag*{\IEEEQEDhere}
\]
\end{IEEEproof}

\section{Proof of Lemma~\ref{z:prop-fourier-coefficients}}
\label{app:fourier-coefficients}

\begin{IEEEproof}
\emph
Using \eqref{z:eq-Za} in \eqref{z:eq-ch-def} to obtain
\[
 c_h(a)=\lim_{M\to\infty}\sum_{k,k'=-M}^{M}\int_0^1
 \Phi_a(\nu_k)\overline{\Phi_a(\nu_{k'})}\e^{2\pi ih\nu}\dd\nu.
\]
Set \(d=k'-k\) and \(x=\nu_k=k+\tfrac12-\nu\). Then
\begin{equation}
 \begin{aligned}
 c_h(a) &=\!\lim_{M\to\infty}\!
 \sum_{\substack{|k|\leq M\\ |k+d|\leq M}}\!
 \int_{k-\frac{1}{2}}^{k+\frac{1}{2}}\!
 \Phi_a(x)\overline{\Phi_a(x+d)} \e^{-2\pi ih(x-\frac{1}{2})}\dd x\\
 &=(-1)^h \sum_{d\in\Z}\int_{\mathbb R}
 \Phi_a(x)\overline{\Phi_a(x+d)}\e^{-2\pi ihx}\dd x.
 \end{aligned}
 \label{z:eq-periodic-correlation}
\end{equation}

By \eqref{m:finite-kernel},
\[
 \begin{aligned}
 &\int_{\mathbb R}\Phi_a(x)\overline{\Phi_a(x+d)}
                     \e^{-2\pi ihx}\dd x\\
 =&a\e^{-i\pi ad^2}\int_{\mathbb R}
       \sinc(ax)\sinc(a(x+d)) \e^{-2\pi i(h+ad)x}\dd x.
 \end{aligned}
\]
Noting the following Fourier correspondence:
\begin{equation}
 \begin{aligned}
 \sinc(ax)\e^{-2\pi i(h+ad)x}
 &\longleftrightarrow a^{-1}\ind_I(\eta+h+ad),\\
 \sinc(a(x+d))
 &\longleftrightarrow a^{-1}\e^{2\pi id\eta}\ind_I(\eta),
 \end{aligned}
 \label{z:eq-sinc-transform}
\end{equation}
where \(I=[-a/2,a/2]\), then Plancherel's theorem gives
\begin{equation}
 \begin{aligned}
 &\int_{\mathbb R}\Phi_a(x)\overline{\Phi_a(x+d)}
                     \e^{-2\pi ihx}\dd x\\
 =&\ a\e^{-i\pi ad^2}\int_{\mathbb R}
       a^{-2}\ind_I(\eta+h+ad)\ind_I(\eta) \e^{-2\pi id\eta}\dd\eta\\
 =&\ \frac{\e^{-i\pi ad^2}}a
       \int_{I\cap(I-h-ad)}\e^{-2\pi id\eta}\dd\eta.
 \end{aligned}
 \label{z:eq-overlap-integral}
\end{equation}
The interval \(I\cap(I-h-ad)\) is nonempty exactly when
\(|h+ad|<a\). Write \(h/a=\kappa_h+\beta\), \(0\leq\beta<1\).
Then \(|\kappa_h+\beta+d|<1\) leaves only \(d=-\kappa_h,-\kappa_h-1\), with
\begin{equation}
 I\cap(I-h-ad)=
 \begin{cases}
 [-a/2,a/2-a\beta],&d=-\kappa_h,\\
 [a/2-a\beta,a/2],&d=-\kappa_h-1,
 \end{cases}
 \label{z:eq-two-overlap-intervals}
\end{equation}
where the second interval vanishes if \(\beta=0\). Integrating \eqref{z:eq-overlap-integral} over these intervals and summing up their contributions in \eqref{z:eq-periodic-correlation} gives \eqref{z:eq-ch-formula}.
\end{IEEEproof}

\bibliographystyle{IEEEtran}
\AddToHook{env/thebibliography/begin}{\interlinepenalty=10000}
\bibliography{references}

\begin{thebibliography}{10}
\providecommand{\url}[1]{#1}
\csname url@samestyle\endcsname
\providecommand{\newblock}{\relax}
\providecommand{\bibinfo}[2]{#2}
\providecommand{\BIBentrySTDinterwordspacing}{\spaceskip=0pt\relax}
\providecommand{\BIBentryALTinterwordstretchfactor}{4}
\providecommand{\BIBentryALTinterwordspacing}{\spaceskip=\fontdimen2\font plus
\BIBentryALTinterwordstretchfactor\fontdimen3\font minus
  \fontdimen4\font\relax}
\providecommand{\BIBforeignlanguage}[2]{{%
\expandafter\ifx\csname l@#1\endcsname\relax
\typeout{** WARNING: IEEEtran.bst: No hyphenation pattern has been}%
\typeout{** loaded for the language `#1'. Using the pattern for}%
\typeout{** the default language instead.}%
\else
\language=\csname l@#1\endcsname
\fi
#2}}
\providecommand{\BIBdecl}{\relax}
\BIBdecl

\bibitem{Ksienski1960}
A.~Ksienski, ``Maximally flat and quasi-smooth sector beams,'' \emph{IRE Trans.
  Antennas Propag.}, vol.~8, no.~5, pp. 476--484, Sep. 1960, {doi}:
  10.1109/TAP.1960.1144881.

\bibitem{OrchardElliottStern1985}
H.~J. Orchard, R.~S. Elliott, and G.~J. Stern, ``Optimising the synthesis of
  shaped beam antenna patterns,'' \emph{IEE Proc. H, Microw. Antennas Propag.},
  vol. 132, no.~1, pp. 63--68, Feb. 1985, {doi}: 10.1049/ip-h-2.1985.0013.

\bibitem{KerceBrownMitchell2007}
J.~C. Kerce, G.~C. Brown, and M.~A. Mitchell, ``Phase-only transmit beam
  broadening for improved radar search performance,'' in \emph{Proc. {IEEE}
  Radar Conf.}, Apr. 2007, pp. 451--456, {doi}: 10.1109/RADAR.2007.374259.

\bibitem{GirnykPetersson2021}
M.~A. Girnyk and S.~O. Petersson, ``Efficient cell-specific beamforming for
  large antenna arrays,'' \emph{IEEE Trans. Commun.}, vol.~69, no.~12, pp.
  8429--8442, 2021.

\bibitem{DuJiang2024}
C.~Du and Y.~Jiang, ``Broad beam designs for broadcast channels,'' \emph{IEEE
  Trans. Signal Process.}, vol.~72, pp. 3819--3833, 2024, {doi}:
  10.1109/TSP.2024.3440353.

\bibitem{XiaoXiaXia2017}
Z.~Xiao, P.~Xia, and X.-G. Xia, ``Codebook design for millimeter-wave channel
  estimation with hybrid precoding structure,'' \emph{IEEE Trans. Wireless
  Commun.}, vol.~16, no.~1, pp. 141--153, Jan. 2017.

\bibitem{XiaGao2016}
X.-G. Xia and X.~Gao, ``A space-time code design for omnidirectional
  transmission in massive {MIMO} systems,'' \emph{IEEE Wireless Commun. Lett.},
  vol.~5, no.~5, pp. 512--515, Oct. 2016, {doi}: 10.1109/LWC.2016.2597231.

\bibitem{LiJiangDuWang2021}
F.~Li, Y.~Jiang, C.~Du, and X.~Wang, ``Construction of {Golay} complementary
  matrices and its applications to {MIMO} omnidirectional transmission,''
  \emph{IEEE Trans. Signal Process.}, vol.~69, pp. 2100--2113, 2021, {doi}:
  10.1109/TSP.2021.3067467.

\bibitem{3GPPTS38300}
{3GPP}, ``{NR}; overall description; stage-2,'' 3rd Generation Partnership
  Project, Technical Specification 38.300, Oct. 2018, v15.3.1, Release 15.
  \href{https://www.etsi.org/deliver/etsi_ts/138300_138399/138300/15.03.01_60/ts_138300v150301p.pdf}{Available
  online}.

\bibitem{GirnykJidhageFaxer2023}
M.~Girnyk, H.~Jidhage, and S.~Fax{\'e}r, ``Broad beamforming technology in {5G}
  massive {MIMO},'' \emph{Ericsson Technol. Rev.}, vol. 2023, no.~10, pp. 2--6,
  Oct. 2023, {doi}: 10.23919/ETR.2023.10313525.

\bibitem{Golay1982}
M.~J.~E. Golay, ``The merit factor of long low autocorrelation binary
  sequences,'' \emph{IEEE Trans. Inf. Theory}, vol.~28, no.~3, pp. 543--549,
  May 1982, {doi}: 10.1109/TIT.1982.1056505.

\bibitem{Schmidt2016Survey}
K.-U. Schmidt, ``Sequences with small correlation,'' \emph{Des. Codes
  Cryptogr.}, vol.~78, no.~1, pp. 237--267, 2016, {doi}:
  10.1007/s10623-015-0154-7.

\bibitem{HeathEtAl2016}
R.~W. Heath, Jr., N.~Gonz{\'a}lez-Prelcic, S.~Rangan, W.~Roh, and A.~M. Sayeed,
  ``An overview of signal processing techniques for millimeter wave {MIMO}
  systems,'' \emph{IEEE J. Sel. Topics Signal Process.}, vol.~10, no.~3, pp.
  436--453, Apr. 2016, {doi}: 10.1109/JSTSP.2016.2523924.

\bibitem{Golay1961}
M.~J.~E. Golay, ``Complementary series,'' \emph{IRE Trans. Inf. Theory},
  vol.~7, no.~2, pp. 82--87, Apr. 1961, {doi}: 10.1109/TIT.1961.1057620.

\bibitem{DuJiangGolay2024}
C.~Du and Y.~Jiang, ``Polyphase {Golay} complementary arrays: {The} possible
  sizes and new constructions,'' \emph{IEEE Trans. Inf. Theory}, vol.~70,
  no.~2, pp. 1422--1435, Feb. 2024, {doi}: 10.1109/TIT.2023.3313574.

\bibitem{CaoThompsonHaas2017}
P.~Cao, J.~S. Thompson, and H.~Haas, ``Constant-modulus shaped beam synthesis
  via convex relaxation,'' \emph{IEEE Antennas Wireless Propag. Lett.},
  vol.~16, pp. 617--620, 2017, {doi}: 10.1109/LAWP.2016.2594141.

\bibitem{TranterEtAl2017}
J.~Tranter, N.~D. Sidiropoulos, X.~Fu, and A.~Swami, ``Fast unit-modulus least
  squares with applications in beamforming,'' \emph{IEEE Trans. Signal
  Process.}, vol.~65, no.~11, pp. 2875--2887, 2017, {doi}:
  10.1109/TSP.2017.2666774.

\bibitem{AroraEtAl2022}
A.~Arora, C.~G. Tsinos, M.~R.~B. Shankar, S.~Chatzinotas, and B.~Ottersten,
  ``Efficient algorithms for constant-modulus analog beamforming,'' \emph{IEEE
  Trans. Signal Process.}, vol.~70, pp. 756--771, 2022, {doi}:
  10.1109/TSP.2021.3094653.

\bibitem{ColemanEtAl2010}
J.~O. Coleman, D.~P. Scholnik, and K.~R. McPhail, ``Phase-only tapers for
  regular planar arrays---{A} heuristic nonlinear-{FM} approach,'' in
  \emph{Proc. {IEEE} Int. Symp. Phased Array Syst. Technol.}, 2010, pp.
  113--120, {doi}: 10.1109/ARRAY.2010.5613383.

\bibitem{FonteneauEtAl2022}
C.~Fonteneau, M.~Crussi{\`e}re, and B.~Jahan, ``A beam broadening method for
  phased arrays in wireless communications,'' \emph{EURASIP J. Wireless Commun.
  Netw.}, vol. 2022, 2022, art.~no.~91, {doi}: 10.1186/s13638-022-02173-9.

\bibitem{AngelettiEtAl2026I}
P.~Angeletti, G.~Toso, and R.~Vitolo, ``Asymptotic phase synthesis by transport
  maps---{Part I}: {Theory} and irrotational linear maps,'' \emph{IEEE Open J.
  Antennas Propag.}, vol.~7, no.~2, pp. 369--393, 2026, {doi}:
  10.1109/OJAP.2025.3646973.

\bibitem{AngelettiEtAl2026II}
------, ``Asymptotic phase synthesis by transport maps---{Part II}: {Optimal}
  transport problem,'' \emph{IEEE Open J. Antennas Propag.}, vol.~7, no.~2, pp.
  394--405, 2026, {doi}: 10.1109/OJAP.2025.3646965.

\bibitem{Littlewood1966}
J.~E. Littlewood, ``On polynomials $\sum^n \pm z^m$, $\sum^n e^{\alpha_m
  i}z^m$, $z=e^{\theta i}$,'' \emph{J. London Math. Soc.}, vol.~41, no.~1, pp.
  367--376, 1966, {doi}: 10.1112/jlms/s1-41.1.367.

\bibitem{HoholdtJensenJustesen1985}
T.~H{\o}holdt, H.~E. Jensen, and J.~Justesen, ``Aperiodic correlations and the
  merit factor of a class of binary sequences,'' \emph{IEEE Trans. Inf.
  Theory}, vol.~31, no.~4, pp. 549--552, Jul. 1985, {doi}:
  10.1109/TIT.1985.1057071.

\bibitem{KatzLeeTrunov2020}
D.~J. Katz, S.~Lee, and S.~A. Trunov, ``{Rudin-Shapiro}-like sequences with
  maximum asymptotic merit factor,'' \emph{IEEE Trans. Inf. Theory}, vol.~66,
  no.~12, pp. 7728--7738, Dec. 2020, {doi}: 10.1109/TIT.2020.3011853.

\bibitem{JensenJensenHoholdt1991}
J.~M. Jensen, H.~E. Jensen, and T.~H{\o}holdt, ``The merit factor of binary
  sequences related to difference sets,'' \emph{IEEE Trans. Inf. Theory},
  vol.~37, no.~3, pp. 617--626, May 1991, {doi}: 10.1109/18.79917.

\bibitem{Katz2013L4}
D.~J. Katz, ``Asymptotic {$L^4$} norm of polynomials derived from characters,''
  \emph{Pacific J. Math.}, vol. 263, no.~2, pp. 373--398, Jun. 2013, {doi}:
  10.2140/pjm.2013.263.373.

\bibitem{BorweinChoiJedwab2004}
P.~Borwein, K.-K.~S. Choi, and J.~Jedwab, ``Binary sequences with merit factor
  greater than 6.34,'' \emph{IEEE Trans. Inf. Theory}, vol.~50, no.~12, pp.
  3234--3249, Dec. 2004, {doi}: 10.1109/TIT.2004.838341.

\bibitem{JedwabKatzSchmidt2013}
J.~Jedwab, D.~J. Katz, and K.-U. Schmidt, ``Advances in the merit factor
  problem for binary sequences,'' \emph{J. Combin. Theory Ser. A}, vol. 120,
  no.~4, pp. 882--906, May 2013, {doi}: 10.1016/j.jcta.2013.01.010.

\bibitem{Baden2011}
J.~M. Baden, ``Efficient optimization of the merit factor of long binary
  sequences,'' \emph{IEEE Trans. Inf. Theory}, vol.~57, no.~12, pp. 8084--8094,
  Dec. 2011, {doi}: 10.1109/TIT.2011.2164778.

\bibitem{KatzRamirez2025}
D.~J. Katz and M.~E. Ramirez, ``Limiting moments of autocorrelation demerit
  factors of binary sequences,'' \emph{IEEE Trans. Inf. Theory}, vol.~71,
  no.~4, pp. 3181--3193, Apr. 2025, {doi}: 10.1109/TIT.2025.3532557.

\bibitem{OpenAI2026Littlewood}
{OpenAI}, ``Asymptotically minimal maxima of real {Littlewood} polynomials,''
  {OpenAI Math Release} preprint, Sep.~23 2026,
  \href{https://github.com/openai/math/blob/main/preprints/Asymptotically-minimal-maxima-of-real-Littlewood-polynomials-September-23-2026/paper.pdf}{Available
  online}.

\bibitem{Frank1963}
R.~L. Frank, ``Polyphase codes with good nonperiodic correlation properties,''
  \emph{IEEE Trans. Inf. Theory}, vol.~9, no.~1, pp. 43--45, Jan. 1963, {doi}:
  10.1109/TIT.1963.1057798.

\bibitem{Chu1972}
D.~C. Chu, ``Polyphase codes with good periodic correlation properties
  ({Corresp.}),'' \emph{IEEE Trans. Inf. Theory}, vol.~18, no.~4, pp. 531--532,
  Jul. 1972, {doi}: 10.1109/TIT.1972.1054840.

\bibitem{Mercer2013}
I.~Mercer, ``Merit factor of {Chu} sequences and best merit factor of polyphase
  sequences,'' \emph{IEEE Trans. Inf. Theory}, vol.~59, no.~9, pp. 6083--6086,
  Sep. 2013, {doi}: 10.1109/TIT.2013.2259537.

\bibitem{Schmidt2013}
K.-U. Schmidt, ``On a problem due to {Littlewood} concerning polynomials with
  unimodular coefficients,'' \emph{J. Fourier Anal. Appl.}, vol.~19, no.~3, pp.
  457--466, 2013, {doi}: 10.1007/s00041-013-9265-6.

\bibitem{PetrolatiAngelettiToso2012}
D.~Petrolati, P.~Angeletti, and G.~Toso, ``New piecewise linear polyphase
  sequences based on a spectral domain synthesis,'' \emph{IEEE Trans. Inf.
  Theory}, vol.~58, no.~7, pp. 4890--4898, Jul. 2012, {doi}:
  10.1109/TIT.2012.2191664.

\bibitem{BorweinFerguson2005}
P.~Borwein and R.~Ferguson, ``Polyphase sequences with low autocorrelation,''
  \emph{IEEE Trans. Inf. Theory}, vol.~51, no.~4, pp. 1564--1567, Apr. 2005,
  {doi}: 10.1109/TIT.2004.842778.

\bibitem{StoicaHeLi2009}
P.~Stoica, H.~He, and J.~Li, ``New algorithms for designing unimodular
  sequences with good correlation properties,'' \emph{IEEE Trans. Signal
  Process.}, vol.~57, no.~4, pp. 1415--1425, 2009, {doi}:
  10.1109/TSP.2009.2012562.

\bibitem{SongBabuPalomar2015}
J.~Song, P.~Babu, and D.~P. Palomar, ``Optimization methods for designing
  sequences with low autocorrelation sidelobes,'' \emph{IEEE Trans. Signal
  Process.}, vol.~63, no.~15, pp. 3998--4009, 2015, {doi}:
  10.1109/TSP.2015.2425808.

\bibitem{Kahane1980}
J.-P. Kahane, ``Sur les polyn\^omes \`a coefficients unimodulaires,''
  \emph{Bull. London Math. Soc.}, vol.~12, no.~5, pp. 321--342, 1980, {doi}:
  10.1112/blms/12.5.321.

\bibitem{BombieriBourgain2009}
E.~Bombieri and J.~Bourgain, ``On {Kahane}'s ultraflat polynomials,'' \emph{J.
  Eur. Math. Soc.}, vol.~11, no.~3, pp. 627--703, 2009, {doi}:
  10.4171/JEMS/163.

\bibitem{Vaidyanathan1987}
P.~P. Vaidyanathan, ``On predicting a band-limited signal based on past sample
  values,'' \emph{Proc. IEEE}, vol.~75, no.~8, pp. 1125--1127, Aug. 1987.

\bibitem{Katznelson2004}
Y.~Katznelson, \emph{An Introduction to Harmonic Analysis}, 3rd~ed.\hskip 1em
  plus 0.5em minus 0.4em\relax Cambridge, U.K.: Cambridge University Press,
  2004, ch.~I, Sec.~6.2, p.~32, eq.~(6.2), {doi}:
  \href{https://doi.org/10.1017/CBO9781139165372}{10.1017/CBO9781139165372}.

\end{thebibliography}
\end{document}